\RequirePackage[final]{graphicx}
\documentclass[aps, superscriptaddress, prl, twocolumn]{revtex4-1}
\usepackage[utf8]{inputenc}
\usepackage{amsmath}
\usepackage{amssymb}
\usepackage{amsfonts}
\usepackage{mathtools}
\usepackage{fixme}
\usepackage{color}
\usepackage{braket}
\usepackage{dsfont}
\usepackage{hyperref}

\usepackage{tikz}
\usetikzlibrary{calc}

\usetikzlibrary{positioning,arrows}
\usetikzlibrary{decorations}
\usetikzlibrary{decorations.pathmorphing,positioning}
\usetikzlibrary{decorations.markings}

\tikzset
{
hole/.style         = { draw = black, postaction = { decorate }, decoration = { markings, mark = at position .55 with { \arrow[black]{ triangle 45} } } },
spinwave_11/.style      = { draw = cyan, postaction = { decorate }, decoration = { markings, mark = at position .5 with { \arrow[cyan]{ triangle 45} } } },
particle_small_arrow/.style         = { draw = black, postaction = { decorate }, decoration = { markings, mark = at position .5 with { \arrow[scale = 0.6, black]{ triangle 45} } } },
spinwave_12/.style      = { draw = cyan, postaction = { decorate }, decoration = { markings, mark = at position .25 with { \arrow[cyan, >=triangle 45]{<} }, mark = at position .75 with { \arrow[cyan, >=triangle 45]{>} } } },
spinwave_21/.style      = { draw = cyan, postaction = { decorate }, decoration = { markings, mark = at position .25 with { \arrow[cyan, >=triangle 45]{>} }, mark = at position .75 with { \arrow[cyan, >=triangle 45]{<} } } },
spinwave_total/.style           = { decorate, decoration = {snake, amplitude = 2pt, segment length = 7pt } }
}

\tikzset{
 ncbar angle/.initial=90,
 ncbar/.style={
 to path=(\tikztostart)
 -- ($(\tikztostart)!#1!\pgfkeysvalueof{/tikz/ncbar angle}:(\tikztotarget)$)
 -- ($(\tikztotarget)!($(\tikztostart)!#1!\pgfkeysvalueof{/tikz/ncbar angle}:(\tikztotarget)$)!\pgfkeysvalueof{/tikz/ncbar angle}:(\tikztostart)$)
 -- (\tikztotarget)
 },
 ncbar/.default=0.5cm,
}

\tikzset{square left brace/.style={ncbar=0.1cm}}
\tikzset{square right brace/.style={ncbar=-0.1cm}}

\definecolor{myred}{RGB}{214,26,70}
\definecolor{myreddark}{RGB}{76,8,38}
\definecolor{myblue}{RGB}{35,106,185}
\definecolor{mybluedark}{RGB}{19,56,99}
\definecolor{mybluebright}{RGB}{225,236,249}

\def\te{{\rm e}}

\def\bd{{\bf d}}
\def\bk{{\bf k}}
\def\bp{{\bf p}}
\def\bq{{\bf q}}
\def\br{{\bf r}}

\def\bK{{\bf K}}

\def\bdelta{{\pmb \delta}}

\def\pa{\partial}
\def\nn{\nonumber}
\def\Re{{ \rm Re }}
\def\Im{{ \rm Im }}

\def\AF{{ \rm AF }}

\begin{document}
\title{Nonequilibrium hole dynamics in antiferromagnets: damped strings and polarons}
\author{K.\ Knakkergaard Nielsen}
\affiliation{Max-Planck Institute for Quantum Optics, Hans-Kopfermann-Str. 1, D-85748 Garching, Germany}
\affiliation{Department of Physics and Astronomy, Aarhus University, Ny Munkegade, 8000 Aarhus C, Denmark}
\author{T.\ Pohl}
\affiliation{Department of Physics and Astronomy, Aarhus University, Ny Munkegade, 8000 Aarhus C, Denmark}
\author{G.\ M.\ Bruun} 
\affiliation{Department of Physics and Astronomy, Aarhus University, Ny Munkegade, 8000 Aarhus C, Denmark}
\affiliation{Shenzhen Institute for Quantum Science and Engineering and Department of Physics, Southern University of Science and Technology, Shenzhen 518055, China}
\date{\today}

\begin{abstract}
We develop a nonperturbative theory for hole dynamics in antiferromagnetic spin lattices, as described by the $t$-$J$ model. This is achieved by generalizing the selfconsistent Born approximation to nonequilibrium systems, making it possible to calculate the full time-dependent many-body wave function. Our approach reveals three distinct dynamical regimes, ultimately leading to the formation of magnetic polarons. Following the initial ballistic stage of the hole dynamics, coherent formation of string excitations gives rise to characteristic oscillations in the hole density. Their damping eventually leaves behind magnetic polarons that undergo ballistic motion with a greatly reduced velocity. The developed theory provides a rigorous framework for understanding nonequilibrium physics of defects in quantum magnets and quantitatively explains recent observations from cold-atom quantum simulations in the strong coupling regime.
\end{abstract}

\maketitle
Understanding the motion of charge carriers in quantum spin environments is of great fundamental significance in condensed matter physics \cite{Brinkman1970,SchmittRink1988,Shraiman1988,Kane1989,Trugman1990}. Recently, this problem is attracting growing interest \cite{Carlstrom2016,Nagy2017,Grusdt2018,Grusdt2018_2,Grusdt2019,Bohrdt2019,Bohrdt2021,Soriano2020,Nielsen2021,Diamantis2021}, driven by the success of quantum simulations with ultracold atoms in optical lattices \cite{Esslinger2010,Gross2017}. In particular, the realization of the Fermi-Hubbard model \cite{2010Esslinger,Boll2016,Cheuk2016b,Mazurenko2017,Hilker2017,Brown2017,Chiu2018,Brown2019,Koepsell2019,Chiu2019,Brown2020a,Vijayan2020,Hartke2020,Brown2020b,Koepsell2021,Ji2021,Gall2021} combined with single-site resolution techniques \cite{Bakr2009,Sherson2010,Haller2015,Yang2021} makes it possible to probe the structure and quantum dynamics of lattice defects on a microscopic level \cite{Koepsell2019,Chiu2019,Ji2021}. The behavior of holes in the Fermi-Hubbard model and the associated formation of magnetic polarons is intimately connected to the physics of high-temperature superconductivity \cite{Emery1987,Schrieffer1988,Dagotto1994}. Their dynamics, furthermore, constitutes a paradigmatic example of a strongly interacting quantum many-body system out of equilibrium, whose rigorous description has remained an open theoretical problem. Quantum Monte-Carlo simulations \cite{Carlstrom2016,Nagy2017} have allowed to study the short-time dynamics at infinite temperature and discovered a crossover from an initial -- ballistic -- quantum walk towards a diffusive regime. Recent cold-atom experiments traced the microscopic motion of holes in a simulated Fermi-Hubbard model \cite{Ji2021}, and instead found a crossover to another ballistic regime with a reduced effective hole velocity. This experimentally indicates the creation of magnetic polarons and raises exciting open questions about the dynamical process of quasiparticle formation \cite{Bohrdt2020} and their emerging transport properties.

\begin{figure}[t!]
\begin{center}
\includegraphics[width=0.95\columnwidth]{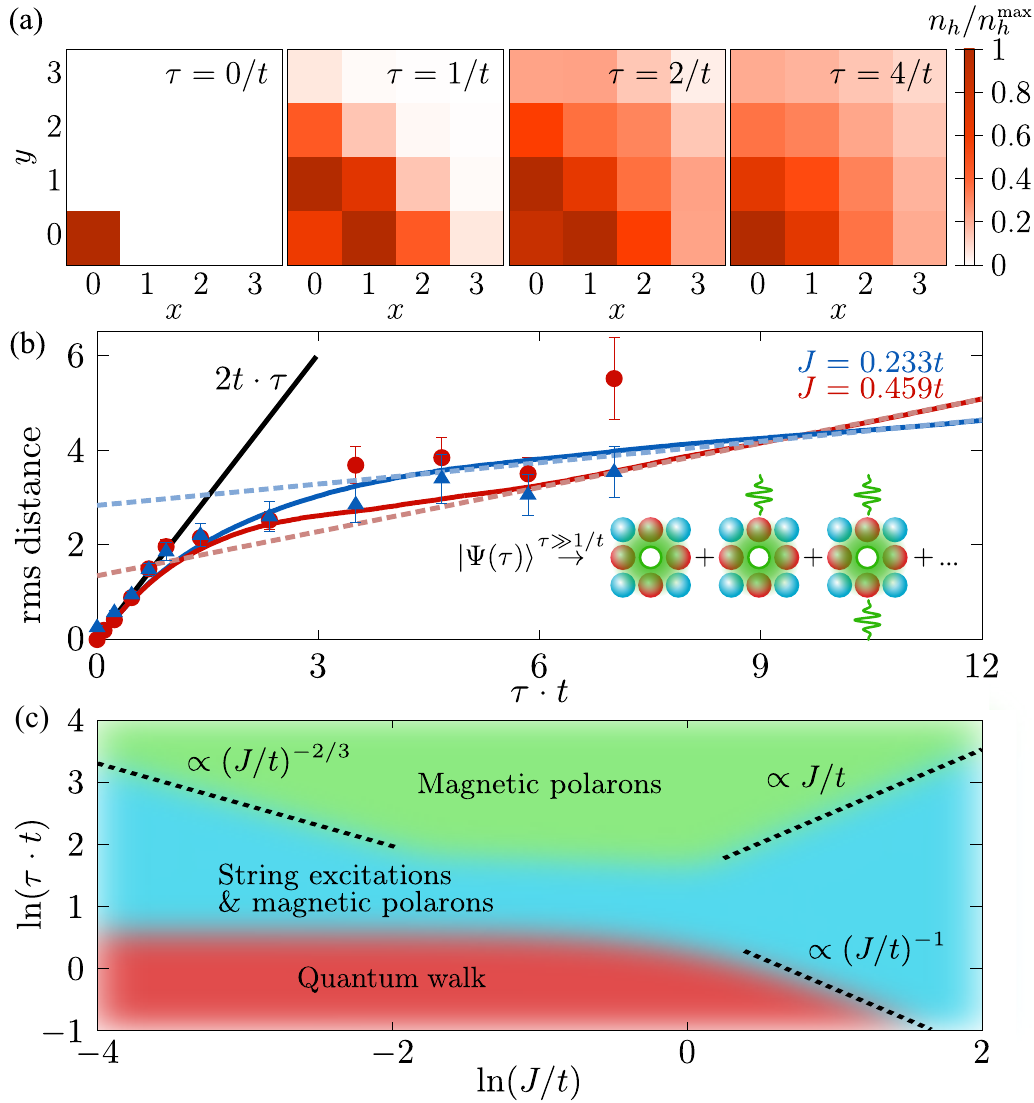}
\end{center}
\vspace{-0.6cm}
\caption{\textbf{Nonequilibrium hole dynamics}. (a) The hole density $n_h({\mathbf d})$ for $J = 0.233t$ and different times normalized to its maximal value $n_{h}^{\rm max}$ is shown in the first quadrant of the lattice, exploiting the $C_4$ rotational symmetry. (b) Root-mean-square distance of the hole as a function of time for indicated interaction strengths compared to experimental results. At long times, the dynamics is determined by ballistic propagation of magnetic polarons (white hole surrounded by red $\sigma=\uparrow$ and blue $\sigma=\downarrow$ fermions), dressed by spin waves (green waves) (inset). (c) We find three distinct dynamical regimes: a quantum walk at short times (red), interfering string excitations at intermediate times (blue), and ballistic transport of magnetic polarons at long times (green).}
\vspace{-0.5cm}
\label{fig.figure_1} 
\end{figure} 

Here, we address this problem and develop a rigorous theoretical framework for the nonequilibrium dynamics of holes in an antiferromagnetic spin lattice. Our starting point is the selfconsistent Born approximation (SCBA) known to be quantitatively accurate under equilibrium conditions \cite{Martinez1991,Liu1991,Diamantis2021}. We generalize this approach to nonequilibrium situations and derive a recursion relation for the time-dependent many-body wave function of the quantum magnet. This makes it possible to describe the complex quantum dynamics of interacting spins as the hole propagates through the underlying antiferromagnetic lattice, accounting for an arbitrary number of spin excitations. Our theory provides a remarkably accurate description of the experimentally observed hole motion at all measured times and strong interactions [see Fig. \ref{fig.figure_1}(b)]. The calculations reveal three dynamical regimes that characterize the dynamical emergence of magnetic polarons from localized lattice defects [Fig. \ref{fig.figure_1}(c)]. The theory predicts a ballistic hole expansion with a universal initial velocity that is independent of the interaction strength [red region in Fig. \ref{fig.figure_1}(c)]. At intermediate times (blue region), the dynamics is characterized by the formation of magnetic polaron states as well as string excitations, in which the hole is confined in the linear potential of flipped spins in its trail \cite{Bulaevskii1968,Brinkman1970,Liu1992,Grusdt2018_2}. Quantum interference between the polarons and the string excitations leads to characteristic oscillations in the hole dynamics consistent with experimental observations \cite{Ji2021}. Ultimately, string excitations are found to dampen at long times, with emergent quasiparticle behavior and ballistic propagation of polarons with a reduced velocity [green region in Fig. \ref{fig.figure_1}(c)]. 

\paragraph{The system.--}
We consider the motion of a single hole in a two-component (spin $\sigma=\uparrow,\downarrow$) Fermi gas in a 2D square lattice. For strong repulsion, the two spins form a quantum antiferromagnet and the system can be described by the $t$-$J$ model \cite{Dagotto1994,Izyumov1997}. Using a slave-fermion representation, this problem can be mapped to the Hamiltonian 
\begin{equation}
\hat{H}= \sum_{\bk} \omega_\bk \hat{b}^\dagger_\bk \hat{b}_\bk+\sum_{\bq, \bk} g(\bq, \bk) [\hat{h}_{\bq + \bk}^\dagger \hat{h}_{\bq} \hat{b}^\dagger_{-\bk} + {\rm H.c.}]
\label{Hamiltonian}
\end{equation}
within linear spin wave theory \cite{SchmittRink1988,Kane1989}. Here, $\hat{b}_\bk^\dagger$ is a bosonic operator creating a spin wave with crystal momentum $\mathbf k$ and energy $\omega_\bk = 2J \sqrt{1 - \gamma_\bk^2}$, where $J>0$ is the antiferromagnetic coupling between neighbouring spins and $\gamma_\bk = [\cos(k_x) + \cos(k_y)] / 2$ is a structure factor, taking the lattice constant to be unity. The second term in Eq. \eqref{Hamiltonian} describes how the motion of holes created by the fermionic operator $\hat h_{\mathbf k}$ makes spin excitations above the antiferromagnetic ground state defined by $\hat{b}_\bk \ket{\AF} = 0$. The associated vertex strength is $g(\bq, \bk) = 4t \cdot (u_\bk \gamma_{\bq + \bk} - v_\bk \gamma_\bq) / \sqrt{N}$ with $N$ the number of lattice sites, $t$ the hopping strength, and $u_\bk = [(1 / \sqrt{1 - \gamma_\bk^2} + 1 ) / 2]^{1/2}$ and $v_\bk = {\rm sgn}(\gamma_\bk) [(1 / \sqrt{1 - \gamma_\bk^2} - 1 ) / 2]^{1/2}$ the coherence factors. The $t$-$J$ model is an effective low-energy description for strong coupling, $J \ll t$, of the Hubbard model realized experimentally \cite{Dagotto1994,Izyumov1997}, and we, therefore, predominantly focus on this regime. 

\paragraph{Time-dependent SCBA.--}
We now describe our time-dependent theory for hole dynamics in the antiferromagnetic lattice. To describe the experimental situation, we initiate a single hole at the site $\bd = {\bf 0}$, i.e. $\ket{\Psi(\tau = 0)} = \hat h^\dagger_{\bf 0} \ket{\AF}=\sum_{\mathbf p}\hat h^\dagger_{\bf p} \ket{\AF}/\sqrt N$, with $\tau$ the variable of time to distinguish it from the hopping $t$. We then have $\ket{\Psi(\tau)} =\sum_{\mathbf p} \ket{\Psi_{\bp}(\tau)}/\sqrt N$, with $i\pa_{\tau} \ket{\Psi_{\bp}(\tau)} = \hat{H}\ket{\Psi_{\bp}(\tau)}$. We write $\ket{\Psi_{\bp}(\tau)} = \ket{\Psi^{\rm R}_{\bp}(\tau)} + \ket{\Psi^{\rm A}_{\bp}(\tau)}$, where $\ket{\Psi^{\rm R}_{\bp}(\tau)} = \exp(-\eta\tau)\theta(\tau) \cdot \ket{\Psi_{\bp}(\tau)}$ and $\ket{\Psi^{\rm A}_{\bp}(\tau)} = \exp(\eta\tau)\theta(-\tau) \cdot \ket{\Psi_{\bp}(\tau)}$ are the retarded and advanced wave functions. Fourier transforming the Schr{\"o}dinger equation then yields \cite{SM}
\begin{equation}
(\omega + i\eta) \ket{\Psi^{\rm R}_{\bp}(\omega)} = i\ket{\Psi_{\bp}(\tau = 0)} + \hat{H}\ket{\Psi^{\rm R}_{\bp}(\omega)}.
\label{eq.equation_of_motion_Psi_R}
\end{equation}
The advanced state is found by $\ket{\Psi^{\rm A}_\bp(\omega)} = [\ket{\Psi^{\rm R}_\bp(\omega)}]^*$. In principle, $\ket{\Psi^{\rm R}_{\bp}(\omega)}$ may be expanded in the number of spin waves. For strong coupling, $J/t\ll1$, however, the corresponding expansion does not truncate in a controlled way, and requires the inclusion of spin waves to \emph{infinite} order.
 
We resolve this problem by generalizing the selfconsistent Born approximation (SCBA) \cite{SchmittRink1988,Kane1989} to nonequilibrium conditions. Note that this approximation yields quantitatively accurate results for the hole Green's function compared to exact diagonalization on small systems \cite{Martinez1991}, and Monte Carlo simulations \cite{Diamantis2021}. In the spirit of the SCBA, we retain only noncrossing terms in the equations of motion for the expansion coefficients of the wave function to obtain a recursion relation for the retarded wave function \cite{SM}
\begin{align}
\ket{\Psi^{\rm R}_\bp(\omega)} =&\, G^{\rm R}(\bp, \omega) \big[i\hat{h}^\dagger_{\bp} \ket{\AF} \nn \\
& + \sum_{\bk} g(\bp, \bk) \cdot \hat{b}^\dagger_{-\bk} \ket{\Psi^{\rm R}_{\bp + \bk}(\omega - \omega_\bk)} \big]. 
\label{eq.Psi_R_frequency_space}
\end{align} 
Here, $G^{\rm R}(\bp, \omega) = [\omega - \Sigma(\bp, \omega) + i\eta]^{-1}$ is the retarded hole Green's function with the SCBA self-energy $\Sigma(\bp, \omega) = \sum_{\bk} g^2(\bp, \bk) G^{\rm R}(\bp + \bk, \omega - \omega_\bk)$ \cite{SchmittRink1988,Kane1989}. The recursive structure in Eq. \eqref{eq.Psi_R_frequency_space} is similar to the SCBA magnetic polaron states \cite{Reiter1994,Nielsen2021}, and allows us to compute the nonequilibrium hole dynamics in an efficient and accurate manner taking an infinite number of spin waves into account. This yields a rigorous generalization of the SCBA to the time-dependent case, and represents the main result of this letter. 

Using the single-site resolution of quantum gas microscopes, one can measure the hole density $n_h(\bd, \tau) = \bra{\Psi(\tau)} \hat{h}^\dagger_{\bd} \hat{h}_\bd \ket{\Psi(\tau)}$ at a given position $\bd$ and time $\tau$ \cite{Ji2021}. Here, we obtain it from
\begin{align}
n_h(\bd, \tau) = \frac{1}{N}\sum_{\bq} \!\te^{-i\bq\cdot\bd} \!\int \frac{{\rm d}\nu}{2\pi} N_h(\bq, \nu), 
\label{eq.hole_density_1}
\end{align}
where $N_h(\bq, \nu) = (2\pi)^{-1}\sum_{\bp}\! \int {\rm d}\omega\, N_h(\bq, \nu; \bp, \omega)$ and $N_h(\bq, \nu; \bp, \omega)=\sum_{\bk}\!\bra{\Psi_{\bp + \bq}(\omega \!+\! \nu)} \hat{h}^\dagger_{\bp+\bq+\bk}\hat{h}_{\bp+\bk}\ket{\Psi_{\bp}(\omega)}$. Using Eq. \eqref{eq.Psi_R_frequency_space}, we derive selfconsistency equations for $N_h$ \cite{SM}, whereby the time-dependent hole density $n_h(\bd, \tau)$ is determined nonperturbatively. We note that the SCBA is a preserving approximation, whereby the total density of holes remains unity, $\sum_\bd n_h(\bd, \tau) = 1$.

\begin{figure}[t!]
\begin{center}
\includegraphics[width=0.95\columnwidth]{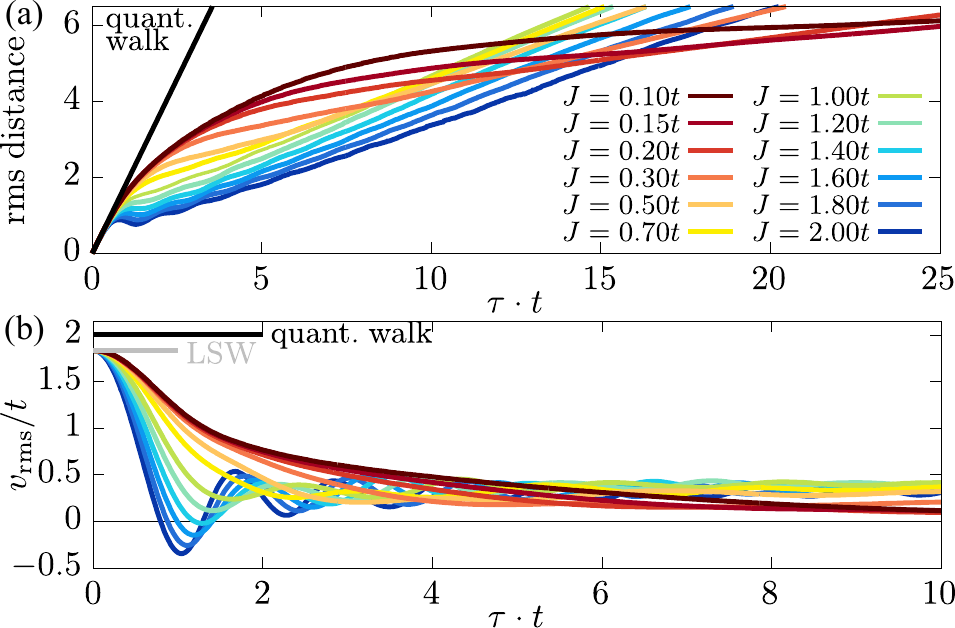}
\end{center}
\vspace{-0.7cm}
\caption{\textbf{Rms distance distance and velocity} (a) Time evolution of the rms distance, $d_{\rm rms}$, for different indicated values of $J / t$. Panel (b) shows the rms velocity $v_{\rm rms} = \pa_\tau d_{\rm rms}$ for the same interaction strengths. The black line corresponds to a free quantum walk with $v_{\rm rms}^0 = 2t$, while the grey line, $v_{\rm rms}^0 \simeq 1.84t$, follows from linear spin wave theory (LSW).}
\vspace{-0.25cm}
\label{fig.rms_dynamics} 
\end{figure} 

\paragraph{Hole dynamics.--}
In Fig. \ref{fig.figure_1}(a), we plot the hole density in the lattice at different times for $J=0.233t$. This illustrates the spatial expansion of the hole following its localized creation at site ${\mathbf d}={\mathbf 0}$. From the density, we determine the root-mean-square (rms) distance, $d_\text{rms}(\tau) = [\sum_{\bd} d^2 \cdot n_h(\bd, \tau)]^{1/2}$, which is compared to experimental measurements \cite{Ji2021} in Fig. \ref{fig.figure_1}(b) for two different coupling strengths. We find a clear crossover between distinct regimes of ballistic expansion with $d=v\tau$ with different velocities $v$. The final expansion velocity is greatly reduced compared to a quantum walk of the hole. For the relevant case of strong coupling, $J = 0.233t$, for which the $t$-$J$ model accurately describes the underlying Fermi-Hubbard Hamiltonian, our theory agrees quantitatively with the experimental data across all timescales. For $J = 0.459t$, the agreement is expectedly not as good, since the mapping from the Fermi-Hubbard Hamiltonian, realized in the experiment \cite{Ji2021}, to the $t$-$J$ model becomes less accurate as $J/t$ increases \cite{Dagotto1994}.

The expansion dynamics is investigated in more detail in Fig. \ref{fig.rms_dynamics}, which displays the rms distance and the associated expansion velocity $v_\text{rms} = \pa_{\tau} d_\text{rms}$ for a range of interaction strengths. This shows that the crossover to the long-time ballistic motion of the polaron slows down with increasing interaction strength $t/J$. Physically, a hole in a lattice with a smaller spin-spin coupling, $J$, will move further away from its initial position before it is affected by the underlying spin order of the quantum magnet. However, a smaller $J$ also implies stronger dressing of the hole by spin waves in its final polaron state. This slows down the long-time ballistic expansion and thereby leads to the nontrivial crossing of the lines in Fig. \ref{fig.rms_dynamics}(a). The calculated dynamics reveals three distinct dynamical regimes: (i) an initial quantum walk of the hole, (ii) a crossover stage driven by string excitations and (iii) the final formation of magnetic polarons. 

\paragraph{(i) Quantum walk.--} 
The results of Figs.~\ref{fig.figure_1} and \ref{fig.rms_dynamics} indicate a universal initial hole dynamics that follows a ballistic expansion with a $J$-independent velocity, $v_{\rm rms}^0$. Indeed, the short-time expansion of the $t$-$J$ model yields a free quantum walk to leading order in $\tau$ with a ballistic expansion velocity $v_{\rm rms}^0=2t$ \cite{Bohrdt_Grusdt_2022,SM}, also observed experimentally~\cite{Ji2021}. The linear spin wave approximation yields a slightly lower velocity of $v_{\rm rms}^0 = 1.84t$ differing from the exact result by only $\sim8$\% [Fig. \ref{fig.rms_dynamics}b]. Expanding our wave function at short times gives $d_\text{rms} \simeq v_{\rm rms}^0 \cdot \tau [1 - c_{3}\cdot (t\cdot \tau)^2]$, where $c_3 = c_{3}^{(0)} + c_{3}^{(J)}\cdot \left(J / t\right)^2$, with interaction-independent coefficients $c_{3}^{(0)}$, $c_{3}^{(J)}$~\cite{SM}. Setting $1 - c_3 (t\tau_s)^2 = 1 / 2$ allows us to define the timescale
\begin{equation}
\tau_s = \frac{1}{t \sqrt{2\left[c_{3}^{(0)} + c_{3}^{(J)}\cdot \left(J / t\right)^2\right]}},
\label{eq.tau_s} 
\end{equation}
after which the initial ballistic behavior breaks down, defining the initial regime shown by the red area in Fig. \ref{fig.figure_1}(c). Consequently, we find $\tau_s$ is proportional to $1/t$ and $1/J$ for strong, $J\ll t$, and weak coupling, $J \gg t$, respectively.

\begin{figure}[t!]
\begin{center}
\includegraphics[width=0.95\columnwidth]{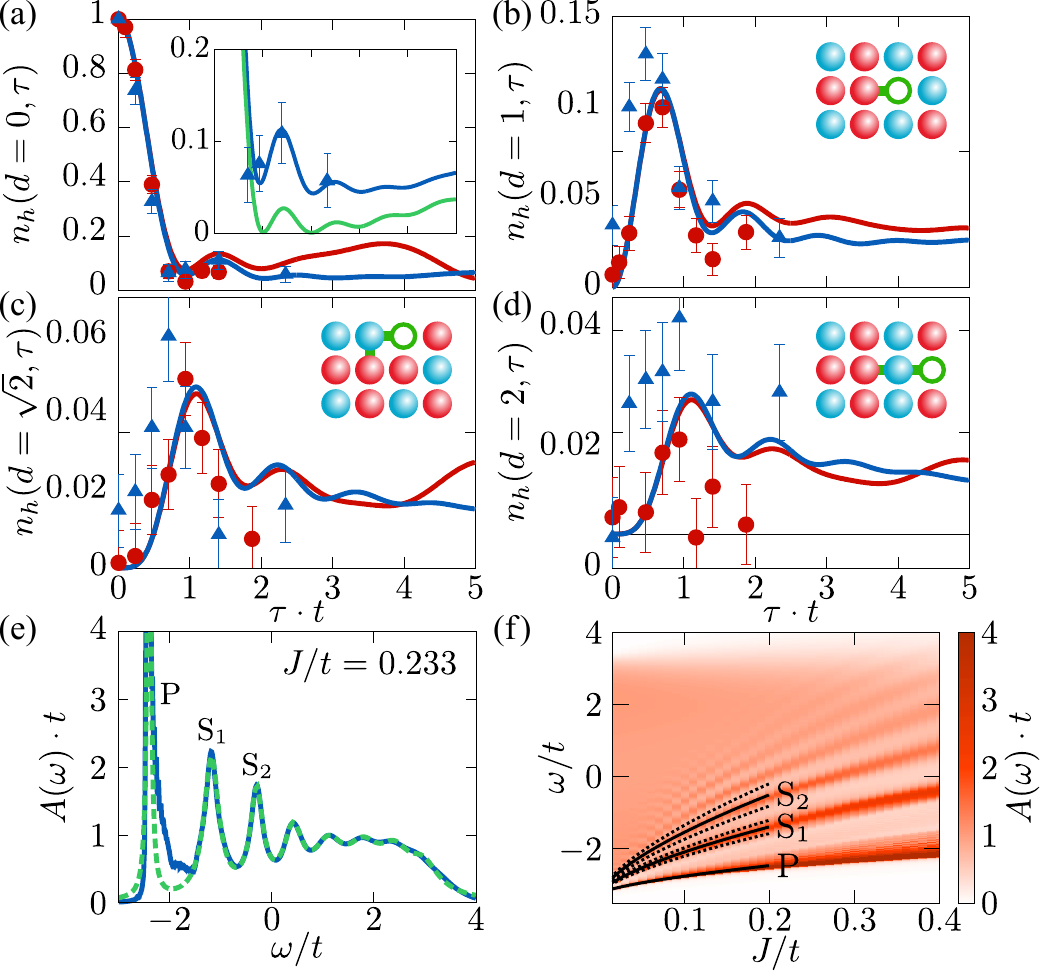}
\end{center}
\vspace{-0.7cm}
\caption{\textbf{Local hole density and string excitations}. (a)--(d) The hole density as a function of time for $J = 0.459t$ (red) and $J = 0.233t$ (blue) for different indicated distances, $d$. The experimental data is shown by red dots for $J = 0.459t$ and by blue triangles for $J = 0.233t$. (e) Total density of states $A(\omega)$ with a multi-Lorentzian fit (dashed green line). In the inset of panel (a), the norm square of the Fourier transform of this fit (green line) is compared to the full solution and experimental data (blue) for $J = 0.233t$. The low-frequency peak of $A(\omega)$ corresponds to magnetic polarons (${\rm P}$). The higher-lying peaks reflect string excitations (${\rm S}_i$). Their energies and spectral widths exhibit a characteristic $(J / t)^{2/3}$ scaling for $J / t \ll 1$, shown by black lines in panel (f).}
\vspace{-0.25cm}
\label{fig.local_hole_density} 
\end{figure}

\paragraph{(ii) Interfering string excitations.--} 
After the initial universal ballistic expansion, the dynamics enters a second regime characterized by oscillations of the hole velocity with a period that increases with the interaction strength $t/J$. This is explored further in Fig. \ref{fig.local_hole_density}, where we show the density of holes around its initial position, ${\bf d} = {\bf 0}$, revealing significant oscillations consistent with the experimental observations \cite{Ji2021}. The agreement between theory and experiment is particularly good at $d = 0$ and $d=1$ [Fig. \ref{fig.local_hole_density}(a) and (b)], while accurate accurate comparisons at larger distances are hindering by a decreasing signal-to-noise ratio \cite{Ji2021}. To understand the origin of these oscillations, we show the total density of states (DOS) $A(\omega) = \sum_\bp A(\bp, \omega)/N$ for $J / t = 0.233$ in Fig. \ref{fig.local_hole_density}(e), and as a function of $J / t$ in Fig. \ref{fig.local_hole_density}(f). Here, $A(\bp, \omega)$ is the hole spectral function \cite{SM}. One clearly observes multiple peaks in the DOS, the lowest corresponding to the emerging magnetic polaron. Figure \ref{fig.local_hole_density}(f) also shows a characteristic $(J / t)^{2/3}$-scaling of the position of the high-frequency peaks. This is consistent with string excitations, which correspond to Airy-like eigenstates of the hole, confined in the linear potential formed by its trail of flipped spins. In the strong coupling limit of the $t$-$J_z$ model, this effective potential has a strength $\propto J_z$ \cite{Bulaevskii1968,Kane1989}. The presence of transverse spin-spin couplings broadens the string excitations \cite{Liu1992,Diamantis2021}, and using a multi-Lorentzian fit [Fig. \ref{fig.local_hole_density}(e)] we find that their spectral widths show the same $(J / t)^{2/3}$-scaling as their energies. 

To see how these excitations contribute to the intermediate hole dynamics, we compute the Fourier transform of the Lorentzian fit, $G(\tau)$, which exclusively captures the contribution from the different peaks. Its norm square, $|G(\tau)|^2$, recovers the oscillatory hole motion as shown in the inset of Fig. \ref{fig.local_hole_density}(a). Indeed, it follows from Eq. \eqref{eq.Psi_R_frequency_space} that $|G(\tau)|^2$ determines the hole density at the initial site ${\bf d}={\bf 0}$ to lowest order. We conclude from Fig. \ref{fig.local_hole_density}(a) that the oscillations arise due to quantum interference between the polaron states and the string excitations. In the strongly interacting regime, $J \ll t$, these oscillations stretch out as the spacing between the associated energies diminishes [Fig. \ref{fig.local_hole_density}(f)]. This interference process defines the second dynamical regime [blue region in Fig. \ref{fig.figure_1}(c)], while the lifetime of the lowest string excitation [marked as ${\text S}_1$ in Fig. \ref{fig.local_hole_density}(f)] determines the dynamical crossover into the final propagation stage. The corresponding transition time scales as $(J / t)^{-2/3}$ and $J / t$ for strong and weak coupling respectively, and is shown by the upper lines in Fig. \ref{fig.figure_1}(c). 

\paragraph{(iii) Magnetic polarons.--} 
Following the damping of string excitations, the remaining superposition of magnetic polaron states, once more, undergoes ballistic expansion, at a greatly reduced velocity, evident both from Figs. \ref{fig.figure_1}(b) and \ref{fig.rms_dynamics}. Indeed, at times longer than the string lifetime, we can write our wave function as \cite{SM}
\begin{align}
&\ket{\Psi(\tau)} \to \, \frac{1}{\sqrt{N}}\sum_\bp \Big[ \sqrt{Z_{\bp}}\te^{-i\varepsilon_\bp \tau} \ket{\Psi_{\bp}^{\rm pol}}+ \nn \\
&\sum_{\bk} g_1(\bp,\bk) \sqrt{Z_{\bp + \bk}} \te^{-i(\varepsilon_{\bp + \bk} + \omega_{\bk})\tau} \hat{b}^\dagger_{-\bk} \ket{\Psi_{\bp + \bk}^{\rm pol}} + \dots \Big],\!
\label{eq.wave_function_time_domain_asymptotics_1}
\end{align}
where $g_1(\bp,\bk) = g(\bp, \bk) {\rm Re}[G^{\rm R}(\bp, \varepsilon_{\bp + \bk} + \omega_\bk)]$. The first term in Eq. \eqref{eq.wave_function_time_domain_asymptotics_1} corresponds to the ballistic propagation of magnetic polarons $\ket{\Psi_{\bp}^{\rm pol}}$ with crystal momentum $\bp$, energy $\varepsilon_\bp$, and quasiparticle residue $Z_\bp$. The second term describes polaron propagation along with a spin wave, and is the first term in a series in the number of spin waves [Fig. \ref{fig.figure_1}(b), inset]. These asymptotics explicitly confirms the dynamical formation of magnetic polarons, indicated by experiments \cite{Ji2021}. 
 
\begin{figure}[t!]
\begin{center}
\includegraphics[width=0.95\columnwidth]{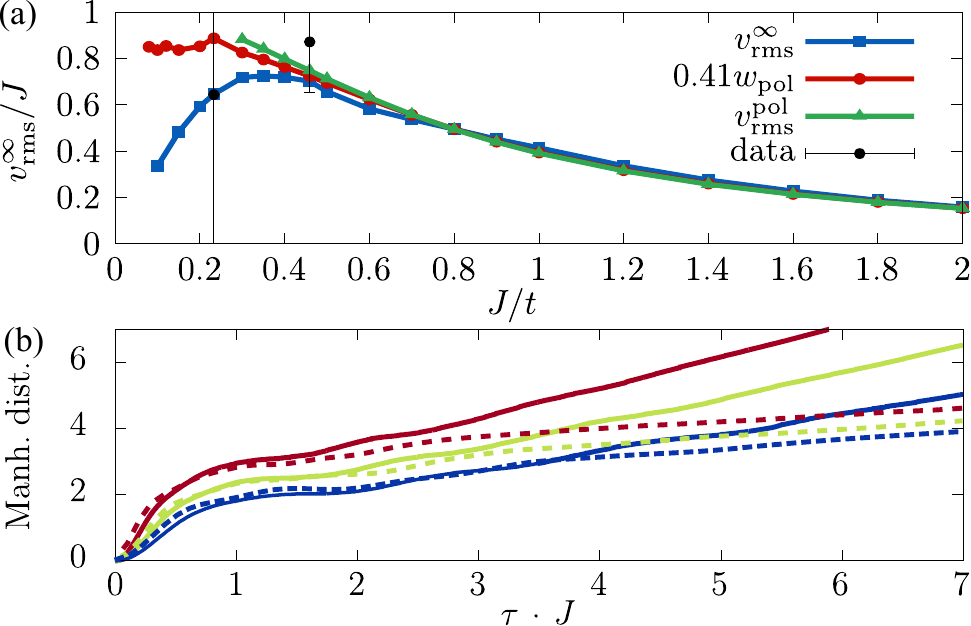}
\end{center}
\vspace{-0.7cm}
\caption{\textbf{Expansion velocity and comparison with MPS simulations}. (a) Long-time expansion velocity $v_{\rm rms}^\infty$ (blue squares), rescaled polaron bandwidth $w_{\rm pol} = \min_\bp \varepsilon_{\bp} - \max_\bp \varepsilon_{\bp}$ (red dots), and average polaronic group velocity $v_{\rm rms}^{\rm pol}$ (green triangles) in units of $J$ as a function of $J / t$. The experimentally observed velocities are shown by the black dots. (b) Manhattan distance for $J/t = 0.35\:({\rm red}),0.5\:({\rm green}),\:{\rm and}\:0.7\:({\rm blue})$ obtained from our dynamical SCBA theory (solid lines) compared to the results of numerical MPS simulations (dashed lines) \cite{Bohrdt2020}.}
\vspace{-0.25cm}
\label{fig.asymptotic_expansion_velocity} 
\end{figure}

In Fig. \ref{fig.asymptotic_expansion_velocity}(a), we show the asymptotic expansion velocity $v_\text{rms}^\infty$. Motivated by the propagation of magnetic polarons evident from Eq.~\eqref{eq.wave_function_time_domain_asymptotics_1}, we also plot the average polaron group velocity $v_{\rm rms}^{\rm pol} = [\sum_\bp (\nabla_\bp\varepsilon_\bp)^2/N]^{1/2}$. In the perturbative limit, the first term in Eq.~\eqref{eq.wave_function_time_domain_asymptotics_1} dominates, and $v_\text{rms}^\infty$ and $v_{\rm rms}^{\rm pol}$ both approach an asymptotic value of $0.41 w_{\rm pol}$ \cite{SM}, evident in Fig.~\ref{fig.asymptotic_expansion_velocity}(a) for $J/t\gtrsim 1$. Below $J / t \simeq 0.4$, however, these two velocities start to deviate significantly. This originates in a qualitative change in the quasiparticle residues, which become very small in certain regions of the Brillouin zone for strong interactions \cite{SM}. As a result, the associated polaron states do not contribute to the long-time dynamics, leading to a sublinear dependency of the expansion velocity on $J$ [Fig. \ref{fig.asymptotic_expansion_velocity}(a)], even though the polaron bandwidth approaches $w_{\rm pol}\sim 2J$ \cite{Kane1989,Dagotto1990}. For very strong coupling $J / t \lesssim 0.02$ \cite{White2001}, it is expected that the ground state of the system develops a growing region of ferromagnetic correlations, the so-called Nagaoka limit \cite{Nagaoka1966}. While we observe indications of this behaviour~\cite{Nielsen2021}, the total spin is conserved under the dynamical evolution, which makes it difficult to observe large ferromagnetic domains in the limit of very small values of $J/t$. The importance of this effect for the motion of holes, hence, remains an interesting topic for future investigations.

In Fig. \ref{fig.asymptotic_expansion_velocity}(b), we compare our results for the Manhattan distance, $\sum_{\bd} (|d_x| + |d_y|)\cdot n_h(\bd,\tau)$, to recent numerical simulations of the $t$-$J$ model using matrix product states (MPS)~\cite{Bohrdt2020}. The good agreement with the numerical results found at short times confirms the accuracy of our approach. At longer times, however, we observe significant deviations, which must be expected due to finite size effects when the Manhattan distance exceeds half the circumference of the $4\times18$ cylindrical lattice simulated in Ref.~\cite{Bohrdt2020}. Reaching large system sizes remains a challenge in numerical simulations, such that the presented theory offers an important approach to explore the dynamics of lattice defects and quasiparticle formation over the complete range of relevant timescales. 

\paragraph{Conclusions.--}
We have developed a nonperturbative approach for computing the nonequilibrium dynamics of holes in Heisenberg antiferromagnets. Our theory provides a quantitative explanation of recent results from cold-atom experiments \cite{Ji2021} at strong interactions. The method yields a complete characterization of the quantum motion of holes and reveals three distinct dynamical regimes that characterize the emergence and evolution of magnetic polarons. It explains observed oscillatory behavior in terms of quantum interference between polarons and string excitations. The presented formalism offers a powerful framework to describe the nonequilibrium quantum dynamics of impurities in strongly interacting lattice models. For example, the method could be utilized to analyze spectroscopic measurements \cite{Bohrdt2018} of holes. Cold-atom experiments also make it possible to probe spin correlations induced by adjacent holes \cite{Koepsell2019,Chiu2019}. Our approach can be used to describe the dynamical buildup of such correlations, and may reveal the dynamics of correlations between two holes, which could provide new insights into pairing and potential mechanisms for high-temperature superconductivity at low doping \cite{Schrieffer1988,Izyumov1997,Shraiman1989,Frenkel1990,Eder1992,Riera1998}. Finally, understanding the impact of temperature on the properties of magnetic polarons \cite{Igarashi1993} remains a challenging open problem, which we hope to explore in the future. 

\begin{acknowledgments}
The authors thank Martin Lebrat, Muqing Xu, Lev Kendrick, Anant Kale and Markus Greiner for sharing experimental data and comments on the manuscript. We thank Annabelle Bohrdt and Fabian Grusdt for valuable feedback on the manuscript. We appreciate helpful discussions with Jens Havgaard Nyhegn. This work has been supported by the Danish National Research Foundation through the Center of Excellence “CCQ” (Grant agreement no.: DNRF156) and by the Carlsberg Foundation through a Carlsberg Internationalisation Fellowship. 

\end{acknowledgments}

\bibliographystyle{apsrev4-1}
\bibliography{ref_magnetic_polaron}

\end{document}


\title{Supplemental Material \\ Nonequilibrium hole dynamics in antiferromagnets: damped strings and polarons}

\maketitle

\beginsupplement
\tableofcontents

\section{Nonequilibrium wave function} \label{sec.non_eq_wave_function}
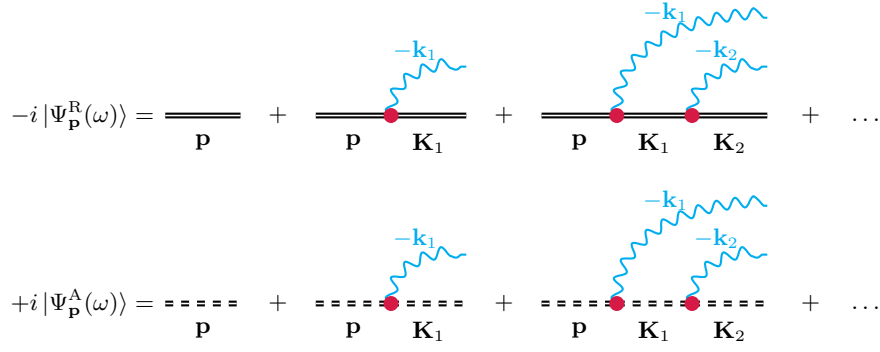
\begin{figure}[t!]
\center
\begin{tikzpicture}[node distance=0.5cm and 1.5cm]
\coordinate[label =center:{$-i\ket{\Psi^{\rm R}_\bp(\omega)} = $}] (a1);
\coordinate[right = 0.6cm of a1] (b1);

\coordinate[right = 0.5cm of b1] (a2);
\coordinate[right = 1.0cm of a2] (b2);
\coordinate[right = 0.5cm of b2, label=center:{$+$}] (c2);

\draw[thick, double] (a2) -- node[label = below:{$\bp$}] {} (b2);

\coordinate[right = 0.5cm of c2] (a3);
\coordinate[right = 1.0cm of a3] (b3);
\coordinate[right = 1.0cm of b3] (c3);
\coordinate[above = 0.65cm of c3] (c3_high);

\coordinate[right = 0.5cm of c3, label=center:{$+$}] (d3);

\draw[thick, double] (a3) --  node[label = below:{$\bp$}] {} (b3);
\draw[thick, double] (b3) -- node[label = below:{$\bK_1$}] {} (c3);
\draw[thick, spinwave_total,cyan] (b3) to[out=90,in=180] node[label = above:{$-\bk_1$}] {}  (c3_high);

\foreach \n in {b3}
  \node at (\n)[circle, fill, inner sep = 2pt, color = myred]{};

\coordinate[right = 0.5cm of d3] (a4);
\coordinate[right = 1.0cm of a4] (b4);
\coordinate[right = 1.0cm of b4] (c4);
\coordinate[right = 1.0cm of c4] (d4);
\coordinate[above = 0.65cm of d4] (d4_high_1);
\coordinate[above = 0.65cm of d4_high_1] (d4_high_2);

\coordinate[right = 1cm of d4, label=center:{$+ \;\;\;\; \dots$}] (e4);

\draw[thick, double] (a4) -- node[label = below:{$\bp$}] {}  (b4);
\draw[thick, double] (b4) -- node[label = below:{$\bK_1$}] {} (c4);
\draw[thick, double] (c4) -- node[label = below:{$\bK_2$}] {} (d4);
\draw[thick, spinwave_total,cyan] (b4) to[out=90,in=180] node[label = above:{$-\bk_1$}] {} (d4_high_2);
\draw[thick, spinwave_total,cyan] (c4) to[out=90,in=180] node[label = above:{$-\bk_2$}] {} (d4_high_1);

\foreach \n in {b4,c4}
  \node at (\n)[circle, fill, inner sep = 2pt, color = myred]{};

\coordinate[below = 2.5cm of a1, label =center:{$+i\ket{\Psi^{\rm A}_\bp(\omega)} = $}] (a1);
\coordinate[right = 0.6cm of a1] (b1);

\coordinate[right = 0.5cm of b1] (a2);
\coordinate[right = 1.0cm of a2] (b2);
\coordinate[right = 0.5cm of b2, label=center:{$+$}] (c2);

\draw[thick, double, dashed] (a2) -- node[label = below:{$\bp$}] {} (b2);

\coordinate[right = 0.5cm of c2] (a3);
\coordinate[right = 1.0cm of a3] (b3);
\coordinate[right = 1.0cm of b3] (c3);
\coordinate[above = 0.65cm of c3] (c3_high);

\coordinate[right = 0.5cm of c3, label=center:{$+$}] (d3);

\draw[thick, double, dashed] (a3) --  node[label = below:{$\bp$}] {} (b3);
\draw[thick, double, dashed] (b3) -- node[label = below:{$\bK_1$}] {} (c3);
\draw[thick, spinwave_total,cyan] (b3) to[out=90,in=180] node[label = above:{$-\bk_1$}] {}  (c3_high);

\foreach \n in {b3}
  \node at (\n)[circle, fill, inner sep = 2pt, color = myred]{};

\coordinate[right = 0.5cm of d3] (a4);
\coordinate[right = 1.0cm of a4] (b4);
\coordinate[right = 1.0cm of b4] (c4);
\coordinate[right = 1.0cm of c4] (d4);
\coordinate[above = 0.65cm of d4] (d4_high_1);
\coordinate[above = 0.65cm of d4_high_1] (d4_high_2);

\coordinate[right = 1cm of d4, label=center:{$+ \;\;\;\; \dots$}] (e4);

\draw[thick, double, dashed] (a4) -- node[label = below:{$\bp$}] {}  (b4);
\draw[thick, double, dashed] (b4) -- node[label = below:{$\bK_1$}] {} (c4);
\draw[thick, double, dashed] (c4) -- node[label = below:{$\bK_2$}] {} (d4);
\draw[thick, spinwave_total,cyan] (b4) to[out=90,in=180] node[label = above:{$-\bk_1$}] {} (d4_high_2);
\draw[thick, spinwave_total,cyan] (c4) to[out=90,in=180] node[label = above:{$-\bk_2$}] {} (d4_high_1);

\foreach \n in {b4,c4}
  \node at (\n)[circle, fill, inner sep = 2pt, color = myred]{};
\end{tikzpicture}
\caption{Diagrammatic representation of nonequilibrium wave functions in frequency space. Green's functions labelled with $\bK_n$ has a shifted frequency of: $\omega - \sum_{j=1}^n \omega_{\bk_j}$. }
\label{fig.diagrammatic_wave_functions}
\end{figure}

In this section, we generalize the self-consistent Born approximation to derive the full nonequilibrium dynamics of the state initialized as a single hole in the crystal momentum state $\bp$
\begin{equation}
\ket{\Psi_\bp(\tau = 0)} = \hat{h}^\dagger_\bp \ket{\AF}.
\label{eq.Psi_p_initial}
\end{equation}
We use $\tau$ as time variable to distinguish it from the hopping amplitude $t$. It is beneficial to split the state into a retarded and advanced part, according to
\begin{gather}
\ket{\Psi_{\bp}(\tau)} = \ket{\Psi^{\rm R}_{\bp}(\tau)} + \ket{\Psi^{\rm A}_{\bp}(\tau)} = \te^{-\eta|\tau|}\left(\theta(\tau)\ket{\Psi_{\bp}(\tau)} + \theta(-\tau)\ket{\Psi_{\bp}(\tau)}\right), 
\end{gather}
in which $\eta > 0$ is a positive infinitesimal that regularizes the Fourier transformation to frequency space. The resulting equation for the retarded wave function in frequency space is, hereby, 
\begin{align}
(\omega + i\eta) \ket{\Psi^{\rm R}_{\bp}(\omega)} = i\ket{\Psi_{\bp}(\tau = 0)} + \hat{H}\ket{\Psi^{\rm R}_{\bp}(\omega)}, 
\label{eq.Psi_R_Hamiltonian_frequency_space}
\end{align}
with $\ket{\Psi^{\rm R}_{\bp}(\omega)} = \int {\rm d}\tau \te^{+i\omega \tau} \ket{\Psi^{\rm R}_{\bp}(\tau)}$. The equation for $\ket{\Psi^{\rm A}_{\bp}(\omega)}$ can be obtained by complex conjugation. As the hole scatters and emits spin waves, the wave function can be expanded in increasing orders of spin waves, 
\begin{gather}
\ket{\Psi^{\rm R}_{\bp}(\omega)} = R^{(0)}(\bp; \omega) \cdot \hat{h}^\dagger_{\bp} \ket{\AF} + \sum_{\bk} R^{(1)}(\bp, \bk; \omega) \cdot \hat{h}^\dagger_{\bp + \bk} \hat{b}^\dagger_{-\bk} \ket{\AF} + \dots
\label{eq.wave_function_expansion}
\end{gather}
The central approximation of our work is to use the framework of the selfconsistent Born approximation (SCBA). Here, only spin waves absorbed in the same order as they are emitted are taken into account. This is the exact same methodology used in \cite{Reiter1994} to establish a wave function consistent with the SCBA, which we used in our previous paper on the spatial structure of magnetic polarons \cite{Nielsen2021}.  From Eq. \eqref{eq.Psi_R_Hamiltonian_frequency_space}, the equations of motion, thereby, become
\begin{gather}
\left[\omega + i\eta \right] R^{(0)}(\bp; \omega) = i + \sum_{\bk} g(\bp, \bk_1) R^{(1)}(\bp, \bk_1; \omega), \nn \\
\left[\omega + i\eta - \sum_{j=1}^n \omega_{\bk_j}\right]\!\! R^{(n)}(\bp, \left\{\bk_j\right\}_{j=1}^n; \omega) = g(\bK_{n-1}, \bk_n) R^{(n-1)}(\bp, \left\{\bk_j\right\}_{j = 1}^{n-1}; \omega) +\! \sum_{\bk_{n+1}} g(\bK_n, \bk_{n+1})R^{(n+1)}(\bp, \left\{\bk_j\right\}_{j=1}^{n+1}; \omega). 
\label{eq.equations_of_motion_retarded_state}
\end{gather}
We note that from these equations of motion, it follows that the norm of the wave function is preserved, $\braket{\Psi_\bp(\tau)|\Psi_\bp(\tau)} = 1$, in the limit of $\eta \to 0$. Inspired by the result in \cite{Reiter1994}, we propose a similiar recursion relation for the coefficients
\begin{gather}
R^{(n+1)}(\bp, \{\bk_j\}_{j = 1}^{n+1}; \omega) = g(\bK_n, \bk_{n+1}) \cdot G^{{\rm R}}(\bK_{n+1}, \omega - \sum_{j = 1}^{n+1}\omega_{\bk_j}) \cdot R^{(n)}(\bp, \{\bk_j\}_{j = 1}^n; \omega), 
\label{eq.recursion_relation}
\end{gather}
where $\bK_{n} = \bp + \sum_{j = 1}^n \bk_j$, and $\bK_0 = \bp$. This entails the retarded hole Green's function $G^{\rm R}(\bp, \omega) = (\omega - \Sigma(\bp, \omega) + i\eta)^{-1}$. The self-energy is determined by the selfconsistency equation inherent to the SCBA \cite{SchmittRink1988,Kane1989,Martinez1991,Liu1991}
\begin{gather}
\Sigma(\bp, \omega) = \sum_{\bk} \frac{g^2(\bp, \bk)}{\omega - \omega_\bk - \Sigma(\bp + \bk, \omega - \omega_\bk) + i\eta}. 
\label{eq.self_energy}
\end{gather}
Using this recursion relation at order $n+1$ in Eq. \eqref{eq.equations_of_motion_retarded_state}, yields
\begin{gather}
\left[\omega + i\eta - \sum_{j=1}^n \omega_{\bk_j} - \Sigma\left(\bK_n, \omega - \sum_{j=1}^n \omega_{\bk_j}\right)\right] R^{(n)}(\bp, \left\{\bk_j\right\}_{j=1}^n; \omega) = g(\bK_{n-1}, \bk_n) R^{(n-1)}(\bp, \left\{\bk_j\right\}_{j = 1}^{n-1}; \omega), 
\end{gather}
which by rearrangement shows the recursion relation in Eq. \eqref{eq.recursion_relation} for order $n$. Finally, using this relation at order $1$ yields $R^{(0)}(\bp; \omega) = iG^{\rm R}(\bp, \omega)$. In this manner, the SCBA makes it possible to write down a wave function that takes an infinite number of spin waves into account. The explicit form of the coefficients for $n\geq 1$ is
\begin{align}
A^{(n)*}(\bp, \left\{\bk_j\right\}_{j=1}^n; \omega) = R^{(n)}(\bp, \left\{\bk_j\right\}_{j=1}^n; \omega) &= +i G^{\rm R}(\bp, \omega) \cdot \prod_{j=1}^n g(\bK_{j-1}, \bk_j) G^{\rm R}(\bK_j, \omega - \sum_{l = 1}^j \omega_{\bk_l}),
\label{eq.R_n_and_A_n_coefficients}
\end{align}
where $A^{(n)}$ are the coefficients of $\ket{\Psi^{\rm A}_\bp(\omega)}$. In this way, we can also write a recursive equation for the states
\begin{align}
[\ket{\Psi^{\rm A}_\bp(\omega)}]^* = \ket{\Psi^{\rm R}_\bp(\omega)} &= G^{\rm R}(\bp, \omega) \left[+i\hat{h}^\dagger_{\bp} \ket{\AF} + \sum_{\bk} g(\bp, \bk) \cdot \hat{b}^\dagger_{-\bk} \ket{\Psi^{\rm R}_{\bp + \bk}(\omega - \omega_\bk)} \right].
\label{eq.recursive_wave_function}
\end{align}
This recursive structure makes it possible for us to compute the motion of the hole with a high accuracy. It also allows us to formulate a diagrammatic expression for the wave functions as shown in Fig. \ref{fig.diagrammatic_wave_functions}, and corresponds to an extension of the method used in Ref. \cite{Ramsak1998} for the magnetic polaron states. The diagrammatic rules are: \vspace{0.2cm} \\
(1) nth full double line: retarded hole Green's function, $G^{\rm R}\left(\bK_n, \varepsilon_\bp - \sum_{i = 1}^n \omega_{\bk_i}\right)$. \\(2) nth dashed double line: advanced hole Green's function, $G^{\rm A}\left(\bK_n, \varepsilon_\bp - \sum_{i = 1}^n \omega_{\bk_i}\right)$. \\
(3) wavy blue line with momentum $-\bk_i$: spinwave operator $\hat{b}^\dagger_{-\bk_i}$.\\
(4) nth red dot in a given diagram: interaction vertex, $g(\bK_n, \bk_{n+1}) / \sqrt{N}$. \\
(5) nth open-ended double line: hole operator, $h^\dagger_{\bK_n}$.\\
(6) finally, sum over all spinwave momenta, $\bk_i$. 

\section{Hole density dynamics}
In this section, we will use the dynamical crystal momentum states in the previous section to calculate the hole density dynamics. Hence, consider the density of holes
\begin{gather}
n_h(\br, \tau) = \bra{\Psi(\tau)} \hat{h}^\dagger_{\br} \hat{h}_\br \ket{\Psi(\tau)}.
\label{eq.hole_density_1}
\end{gather}
Here, the wave function $\ket{\Psi(\tau)}$ is initially given by a hole localized at the central site $\ket{\Psi(\tau)} = \hat{h}^\dagger_{\bd = {\bf 0}} \ket{\AF} = \frac{1}{\sqrt{N}} \sum_{\bp} \hat{h}^\dagger_{\bp} \ket{\AF}$. By decomposing into crystal momentum and frequency components, we can, hereby, write
\begin{gather}
n_h(\bd, \tau) = \frac{1}{N}\sum_{\bq} \te^{-i\bq\cdot\bd} N_{h}(\bq, \tau), 
\label{eq.hole_density_2}
\end{gather}
with $N_{h}(\bq, \tau) = (2\pi)^{-2} \int {\rm d}\nu\, \te^{i\nu \tau} \int {\rm d}\omega \sum_{\bp} N_h(\bq, \nu; \bp, \omega) / N$, and 
\begin{gather}
N_h(\bq, \nu; \bp, \omega) = \sum_{\bk} \bra{\Psi_{\bp + \bq}(\omega + \nu)} \hat{h}^\dagger_{\bp+\bq+\bk}\hat{h}_{\bp+\bk}\ket{\Psi_{\bp}(\omega)}.
\end{gather}
Furthermore, the splitting into the retarded and advanced state, $\ket{\Psi(\omega)} = \ket{\Psi^{\rm R}(\omega)} + \ket{\Psi^{\rm A}(\omega)}$, means that we can write 
\begin{gather}
N_h(\bq, \nu; \bp, \omega) = 2{\rm Re}[N^{\rm RR}_h(\bq, \nu; \bp, \omega) + N^{\rm RA}_h(\bq, \nu; \bp, \omega)],
\label{eq.N_h_and_N_h_RR_RA}
\end{gather}
with $N^{\rm IL}(\bq, \nu; \bp, \omega) = \sum_{\bk} \bra{\Psi^{\rm I}_{\bp + \bq}(\omega + \nu)} \hat{h}^\dagger_{\bp+\bq+\bk}\hat{h}_{\bp+\bk}\ket{\Psi^{\rm L}_{\bp}(\omega)}$, and I, L $=$ R, A. Using the recursive expressions in Eq. \eqref{eq.recursive_wave_function}, we then get
\begin{align}
&N^{\rm RR}_h(\bq, \nu; \bp, \omega) =\, G^{\rm R*}(\bp + \bq, \omega + \nu) G^{\rm R}(\bp, \omega) \nn \\
&\, \cdot \left[1 + \sum_{\bk_1, \bk_2} g(\bp + \bq, \bk_1) g(\bp, \bk_2) \bra{\Psi^{\rm R}_{\bp + \bq + \bk_1}(\omega + \nu - \omega_{\bk_1})} \hat{b}_{-\bk_1} \sum_{\bk_3}\hat{h}^\dagger_{\bp+\bq+\bk_3}\hat{h}_{\bp+\bk_3} \; \hat{b}^\dagger_{-\bk_2}\ket{\Psi_{\bp + \bk_2}(\omega - \omega_{\bk_2})} \right] \nn \\
&\overset{\rm SCBA}{=} G^{\rm R*}(\bp + \bq, \omega + \nu) G^{\rm R}(\bp, \omega) \nn \\
&\, \cdot \left[1 + \sum_{\bk} g(\bp + \bq, \bk) g(\bp, \bk) \bra{\Psi^{\rm R}_{\bp + \bq + \bk}(\omega + \nu - \omega_{\bk})} \sum_{\bk_3}\hat{h}^\dagger_{\bp+\bq+\bk_3}\hat{h}_{\bp+\bk_3} \ket{\Psi_{\bp + \bk}(\omega - \omega_{\bk})} \right] \nn \\
&=  G^{\rm R*}(\bp + \bq, \omega + \nu) G^{\rm R}(\bp, \omega) \cdot \left[1 + \sum_{\bk} g(\bp + \bq, \bk) g(\bp, \bk) N^{\rm RR}_h(\bq, \nu; \bp + \bk, \omega - \omega_\bk) \right]. 
\label{eq.N_RR_self_consistency_equation}
\end{align}
In the second equal sign, we use that spin waves are annihilated in the same order as they are created within the SCBA. Effectively, only one order of the Wick contractions is retained. In a similar manner, it follows that
\begin{gather}
N^{\rm RA}_h(\bq, \nu; \bp, \omega) = -G^{\rm R*}(\bp + \bq, \omega + \nu) G^{\rm A}(\bp, \omega) \cdot \left[1 - \sum_{\bk} g(\bp + \bq, \bk) g(\bp, \bk) N^{\rm RA}_h(\bq, \nu; \bp + \bk, \omega - \omega_\bk) \right]. 
\label{eq.N_RA_self_consistency_equation}
\end{gather}
In this manner, we can iteratively solve for the functions $N^{\rm RR}_h(\bq, \nu; \bp, \omega)$ and $N^{\rm RA}_h(\bq, \nu; \bp, \omega)$, plug these into Eq. \eqref{eq.N_h_and_N_h_RR_RA} to calculate $N_h(\bq, \nu; \bp, \omega)$, which can finally be used in Eq. \eqref{eq.hole_density_2} to calculate the hole density dynamics. 

\section{Short-time dynamics}
In this section, we calculate the short-time dynamics of the hole density dynamics to $\mathcal{O}[(t\cdot \tau)^4]$. To do so, we need to take two spin wave excitation into account
\begin{align}
\ket{\Psi_\bp(\tau)} =&\, C^{(0)}(\bp; \tau) \cdot \hat{h}^\dagger_\bp\ket{\AF} + \sum_{\bk} C^{(1)}(\bp,\bk; \tau) \cdot \hat{h}^\dagger_{\bp + \bk} \hat{b}^\dagger_{-\bk} \ket{\AF} + \sum_{\bk,\bq} C^{(2)}(\bp,\bk,\bq; \tau) \cdot \hat{h}^\dagger_{\bp + \bk + \bq} \hat{b}^\dagger_{-\bq}\hat{b}^\dagger_{-\bk} \ket{\AF}
\label{eq.wave_function_time_domain_4th_order}
\end{align}
Using the Schr{\"o}dinger equation, $i\pa_{\tau}\ket{\Psi_\bp(\tau)} = H \ket{\Psi_\bp(\tau)}$, we get
\begin{align}
i\pa_{\tau} C^{(0)}(\bp; \tau) &= \sum_{\bk}g(\bp, \bk) C^{(1)}(\bp, \bk; \tau), \nn \\
i\pa_{\tau} C^{(1)}(\bp, \bk; \tau) &= \omega_{\bk} C^{(1)}(\bp, \bk; \tau) + g(\bp, \bk)C^{(0)}(\bp; \tau) + \sum_{\bq} g(\bp + \bk,\bq) C^{(2)}(\bp,\bk,\bq;\tau), \nn \\
i\pa_{\tau} C^{(2)}(\bp, \bk; \tau) &= (\omega_{\bk} + \omega_{\bq}) C^{(2)}(\bp, \bk; \tau) + g(\bp + \bk, \bq)C^{(1)}(\bp,\bk; \tau) 
\label{eq.EOM_short_times}
\end{align}
To get an expansion in powers of $t\cdot\tau$, we write
\begin{align}
C^{(0)}(\bp; \tau) &= 1 + C^{(0)}_2(\bp) \cdot \tau^2 + C^{(0)}_3(\bp) \cdot \tau^3 + C^{(0)}_4(\bp) \cdot \tau^4, \nn \\
C^{(1)}(\bp, \bk; \tau) &= C^{(1)}_1(\bp, \bk) \cdot \tau + C^{(1)}_2(\bp, \bk) \cdot \tau^2 + C^{(1)}_3(\bp, \bk) \cdot \tau^3, \nn \\
C^{(2)}(\bp, \bk, \bq; \tau) &= C^{(2)}_2(\bp, \bk, \bq) \cdot \tau^2.
\label{eq.C_0_1_2_short_time}
\end{align}
Generally, at short times, we have $C^{(n)} \propto \tau^n$. The above expansion is enough to get densities that are accurate to fourth order in $\tau$. Looking at the equation of motion for $C^{(0)}(\bp; \tau)$, it furthermore follows that the linear term must be absent, $C^{(0)}_1 = 0$. Inserting Eq. \eqref{eq.C_0_1_2_short_time} into Eq. \eqref{eq.EOM_short_times} and collecting like powers of $\tau$, we get
\begin{align}
C^{(2)}_2(\bp, \bk) &=  -\frac{1}{2} g(\bp,\bk) g(\bp + \bk, \bq).
\label{eq.C_2_j}
\end{align}
for the second order coefficient,
\begin{align}
C^{(1)}_1(\bp, \bk) = -i g(\bp,\bk), \hspace{0.5cm} C^{(1)}_2(\bp, \bk) = -\frac{\omega_\bk}{2} g(\bp,\bk), \hspace{0.5cm} C^{(1)}_3(\bp, \bk) = \frac{i}{6} g(\bp,\bk) \left[ \omega_\bk^2 + g^2_{\rm tot}(\bp) + g^2_{\rm tot}(\bp + \bk)  \right],
\label{eq.C_1_j}
\end{align}
for the first order coefficients, and finally
\begin{align}
C^{(0)}_2(\bp) &= -\frac{1}{2}\sum_{\bk} g^2(\bp,\bk) = -\frac{1}{2}g^2_{\rm tot}(\bp), \hspace{0.5cm} C^{(0)}_3(\bp) = -\frac{i}{3} \sum_{\bk} g(\bp,\bk) C^{(1)}_2(\bp,\bk) = \frac{i}{6}\sum_\bk g^2(\bp,\bk) \cdot \omega_\bk, \nn \\
C^{(0)}_4(\bp) &= -\frac{i}{4} \sum_{\bk} g(\bp,\bk) C^{(1)}_3(\bp,\bk) = \frac{1}{24}\sum_\bk g^2(\bp,\bk)\left[ \omega_\bk^2 + g^2_{\rm tot}(\bp) + g^2_{\rm tot}(\bp + \bk) \right].
\label{eq.C_0_j}
\end{align}
for the zeroth order coefficients. We are now ready to calculate the density to order $(t\cdot \tau)^4$. We utilize that the density may be calculated as an infinite series
\begin{equation}
n_h(\bd, \tau) = \sum_{j=0}^\infty n_h^{(j)}(\bd, \tau).
\label{eq.density_expansion}
\end{equation}
Here, $n_h^{(j)}$ contains the density contribution from the $j$'th order of the state $\ket{\Psi(\tau)}$. The terms are explicitly, 
\begin{align}
n_h^{(0)}(\bd, \tau) &= \left| \bra{\AF} \hat{h}_\bd \ket{\Psi(\tau)} \right|^2 =  \left| \frac{1}{N} \sum_{\bp} \te^{ i\bp\cdot\bd} C^{(0)}(\bp; \tau)\right|^2, \nn \\
n_h^{(1)}(\bd, \tau) &= \sum_{\bk} \left| \bra{\AF} \hat{b}_{-\bk} \hat{h}_\bd \ket{\Psi(\tau)} \right|^2 = \sum_{\bk} \left| \frac{1}{N} \sum_{\bp} \te^{ i\bp\cdot\bd} C^{(1)}(\bp, \bk; \tau)\right|^2, \nn \\
n_h^{(2)}(\bd, \tau) &= \sum_{\bk,\bq} \left| \bra{\AF} \hat{b}_{-\bk}\hat{b}_{-\bq} \hat{h}_\bd \ket{\Psi(\tau)} \right|^2 = \sum_{\bk,\bq} \left| \frac{1}{N} \sum_{\bp} \te^{ i\bp\cdot\bd} C^{(2)}(\bp, \bk, \bq; \tau)\right|^2,
\label{eq.density_expansion_explicit}
\end{align}
and so on for higher orders. Defining $C^{(n)}_j(\bd, \dots) = N^{-1}\sum_\bp \te^{i\bp\cdot\bd} C^{(n)}(\bp, \cdot)$, we may then write
\begin{align}
n_h^{(0)}(\bd, \tau) =&\, \left|C^{(0)}_0(\bd) + C^{(0)}_2(\bd) \cdot \tau^2 + C^{(0)}_3(\bd) \cdot \tau^3 + C^{(0)}_4(\bd) \cdot \tau^4 \right|^2  + \mathcal{O}[(t\cdot \tau)^5] \nn \\
=&\, \left|C^{(0)}_0(\bd)\right|^2 + 2\Re\left[C^{(0)*}_2(\bd)C^{(0)}_0(\bd)\right] \cdot \tau^2 + 2\Re\left[C^{(0)*}_3(\bd)C^{(0)}_0(\bd)\right] \cdot \tau^3 \nn \\
&\, + \left( \left|C^{(0)}_2(\bd)\right|^2 + 2\Re\left[C^{(0)*}_4(\bd)C^{(0)}_0(\bd)\right] \right) \cdot \tau^4 + \mathcal{O}[(t\cdot \tau)^5]\nn \\
=&\; n_{h,0}^{(0)}(\bd) + n_{h,2}^{(0)}(\bd) \cdot (t\cdot \tau)^2 + n_{h,3}^{(0)}(\bd) \cdot (t\cdot \tau)^3 + n_{h,4}^{(0)}(\bd) \cdot (t\cdot \tau)^4 + \mathcal{O}[(t\cdot \tau)^5]. 
\end{align}
Likewise, 
\begin{align}
n_h^{(1)}(\bd, \tau) =&\, \sum_\bk \left|C^{(1)}_1(\bd, \bk) \cdot \tau + C^{(1)}_2(\bd, \bk) \cdot \tau^2 + C^{(1)}_3(\bd, \bk) \cdot \tau^3\right|^2 + \mathcal{O}[(t\cdot \tau)^5] \nn \\
=&\, \sum_\bk \Bigg( \left|C^{(1)}_1(\bd, \bk)\right|^2 \cdot \tau^2 + 2\Re\left[C^{(1)*}_2(\bd, \bk)C^{(1)}_1(\bd, \bk)\right] \cdot \tau^3 \nn \\
&\, + \Big(\left|C^{(1)}_2(\bd, \bk)\right|^2 + 2\Re\left[C^{(1)*}_3(\bd, \bk)C^{(1)}_1(\bd, \bk)\right] \Big) \cdot \tau^4 \Bigg) + \mathcal{O}[(t\cdot \tau)^5]\nn \\
=&\; n_{h,2}^{(1)}(\bd) \cdot (t\cdot \tau)^2 + n_{h,3}^{(1)}(\bd) \cdot (t\cdot \tau)^3 + n_{h,4}^{(1)}(\bd) \cdot (t\cdot \tau)^4 + \mathcal{O}[(t\cdot \tau)^5]. 
\end{align}
From second order in the wave function,
\begin{align}
n_h^{(2)}(\bd, \tau) =&\, \sum_{\bk,\bq} \left|C^{(2)}_2(\bd, \bk,\bq)\right|^2  \cdot \tau^4 + \mathcal{O}[(t\cdot \tau)^5] = n_{h,4}^{(2)}(\bd) \cdot (t\cdot \tau)^4 + \mathcal{O}[(t\cdot \tau)^5]. 
\end{align}
We, finally, reorganize the terms in $n_h(\bd,\tau) \simeq n_h^{(0)}(\bd, \tau) + n_h^{(1)}(\bd, \tau) + n_h^{(2)}(\bd, \tau)$ according to their order in $(t\cdot\tau)$. This gives
\begin{align}
n_h(\bd,\tau) \simeq n_{h,0}^{(0)}(\bd) + \left[ n_{h,2}^{(0)}(\bd) + n_{h,2}^{(1)}(\bd) \right]\cdot(t\cdot \tau)^2 + \left[ n_{h,4}^{(0)}(\bd) + n_{h,4}^{(1)}(\bd) + n_{h,4}^{(2)}(\bd) \right]\cdot(t\cdot \tau)^4. 
\label{eq.n_h_orders_in_tau}
\end{align}
Here, we utilize that the 3rd order terms proportional to $(t\cdot \tau)^3$ all vanish identically, as the enterior of the real values $\Re[\dots]$ turn out to be purely imaginary. We can now write out the explicit terms related to each order. At 0th order in time, we simply have
\begin{align}
n_{h,0}^{(0)}(\bd) &= \left|C^{(0)}_2(\bd)\right|^2 = \delta_{d,0}.
\end{align}
At 2nd order, we get
\begin{align}
n_{h,2}^{(0)}(\bd) &= \frac{2}{t^2}\Re\left[C^{(0)*}_2(\bd)C^{(0)}_0(\bd)\right] = - \delta_{d,0} \cdot \frac{1}{Nt^2} \sum_\bp g^2_{\rm tot}(\bp) \simeq - \delta_{d,0} \cdot 3.37.
\label{eq.n_h_2_0}
\end{align}
The numerical value is in the Heisenberg limit, $\alpha \to 1$, evaluated for a $20\times 20$ lattice. Further, 
\begin{align}
n_{h,2}^{(1)}(\bd) = \frac{1}{t^2} \sum_\bk \left|C^{(1)}_1(\bd, \bk)\right|^2 = \frac{1}{t^2} \sum_\bk \left|\frac{1}{N}\sum_\bp \te^{i\bp\cdot\bd} g(\bp, \bk)\right|^2 = \delta_{d,1} \cdot \frac{1}{N} \sum_\bk \left|u_\bk \te^{-i\bk\cdot\bd} - v_\bk\right|^2 \simeq \delta_{d,1} \cdot 0.843.
\label{eq.n_h_2_1}
\end{align}
Here, we use that $g(\bp, \bk) = zt (u_\bk \gamma_{\bp + \bk} - v_\bk \gamma_\bp) / \sqrt{N}$, and that $\gamma_\bq = \sum_{\bdelta} \te^{i\bq\cdot\bdelta} / z$, to explicitly evaluate the $\bp$ sum. Again, the numerical value is for the Heisenberg limit. The 0th fourth order term evaluates to
\begin{align}
n_{h,4}^{(0)}(\bd) &= \frac{1}{t^4}\left( \left|C^{(0)}_2(\bd)\right|^2 + 2\Re\left[C^{(0)*}_4(\bd)C^{(0)}_0(\bd)\right] \right) \nn \\
&= \frac{1}{t^4}\left( \frac{1}{4} \left|\frac{1}{N} \sum_\bp \te^{i\bp\cdot\bd} g^2_{\rm tot}(\bp)\right|^2 + \frac{\delta_{d,0}}{12} \cdot \frac{1}{N} \sum_{\bp,\bk} g^2(\bp,\bk)\left[\omega_\bk^2 + g^2_{\rm tot}(\bp) + g^2_{\rm tot}(\bp + \bk) \right] \right) \nn \\
&\simeq \delta_{d,0} \left[ 0.917 \cdot \left(\frac{J}{t}\right)^2 + 4.522\right] + \delta_{d,\sqrt{2}} \cdot 0.044 + \delta_{d,2} \cdot 0.014
\label{eq.n_h_4_0}
\end{align}
The 1st fourth order term is, similarly,
\begin{align}
n_{h,4}^{(1)}(\bd) &= \frac{1}{t^4}\sum_\bk \left( \left|C^{(1)}_2(\bd,\bk)\right|^2 + 2\Re\left[C^{(1)*}_3(\bd,\bk)C^{(1)}_1(\bd,\bk)\right] \right) \nn \\
&= \frac{1}{N t^2}\sum_\bk \delta_{d,1} \cdot \left( \frac{\omega_\bk^2}{4} \left|u_\bk \te^{-i\bk\cdot\bd} - v_\bk\right|^2 - \frac{1}{3} \Re\left[(u_\bk \te^{-i\bk\cdot\bd} - v_\bk) \cdot \frac{1}{t\sqrt{N}} \sum_\bp \te^{-i\bp\cdot\bd} g(\bp,\bk) [\omega_\bk^2 + g^2_{\rm tot}(\bp) + g^2_{\rm tot}(\bp + \bk)]\right] \right) \nn \\
&= \frac{1}{N t^2}\sum_\bk \delta_{d,1} \cdot \left( -\frac{\omega_\bk^2}{12} \left|u_\bk \te^{-i\bk\cdot\bd} - v_\bk\right|^2 - \frac{1}{3} \Re\left[(u_\bk \te^{-i\bk\cdot\bd} - v_\bk) \cdot \frac{1}{t\sqrt{N}} \sum_\bp \te^{-i\bp\cdot\bd} g(\bp,\bk) [g^2_{\rm tot}(\bp) + g^2_{\rm tot}(\bp + \bk)]\right] \right) \nn \\
&\simeq \frac{\delta_{d,1}}{4}\left[ -0.917 \left(\frac{J}{t}\right)^2 -6.746\right]
\label{eq.n_h_2_1}
\end{align}
Finally, 
\begin{align}
n_{h,4}^{(2)}(\bd) &= \frac{1}{t^4}\sum_{\bk,\bq} \left|C^{(2)}_2(\bd,\bk)\right|^2 = \frac{1}{4t^4} \sum_{\bk,\bq} \left|\frac{1}{N}\sum_\bp \te^{i\bp\cdot\bd} g(\bp,\bk)g(\bp + \bk,\bq)\right|^2 \nn \\
&= \delta_{d,0} \cdot 0.196 + \delta_{d,\sqrt{2}} \cdot 0.272 + \delta_{d,2} \cdot 0.177
\label{eq.n_h_2_1}
\end{align}
We are now ready to characterize the short-time motion of the hole up to order $(t\cdot\tau)^4$. To this end, we calculate the root-mean-square distance
\begin{align}
d_{\rm rms} &= \left[\sum_{\bd} d^2 n_h(\bd,\tau)\right]^{1/2} \simeq \left[\sum_{\bd} d^2 \! \left(n_{h,0}^{(0)}(\bd) + \left[ n_{h,2}^{(0)}(\bd) + n_{h,2}^{(1)}(\bd) \right](t\cdot \tau)^2 + \left[ n_{h,4}^{(0)}(\bd) + n_{h,4}^{(1)}(\bd) + n_{h,4}^{(2)}(\bd) \right](t\cdot \tau)^4\right)\right]^{1/2}\nn \\
&= v_{\rm rms}^0 \cdot \tau \left[1 - c_3 \cdot (t\cdot \tau)^2\right] + \mathcal{O}[(t\cdot\tau)^5]
\end{align}
Here, 
\begin{equation}
v_{\rm rms}^0 = 2t \cdot \left[\frac{1}{N} \sum_\bk \left|u_\bk \te^{-i\bk\cdot\bd} - v_\bk\right|^2\right]^{1/2} \simeq 1.835t, 
\end{equation}
is the initial expansion velocity for the hole within the linear spin wave approximation. Compared to a free quantum walk of a particle, this has an additional factor of $[\frac{1}{N} \sum_{\bk} \left|u_\bk \te^{-i\bk\cdot\bd} - v_\bk\right|^2 ]^{1/2} \simeq 0.918$ in the Heisenberg limit, $\alpha \to 1$. The hole motion in the full $t$-$J$ model has to have an initial velocity of $2t$. This small deficiency of the linear spin wave theory is a result of an interference between two pathways: (1) it creates a spin excitation at the initial site, $d = 0$, and (2) it absorbs a spin excitation from the neighbouring site it moves to, $d = 1$. In the full $t$-$J$ model these processes are, however, orthogonal as they leave behind either a spin-$\uparrow$ or -$\downarrow$ at the origin. This is shown explicitly in the next Section. Now,
\begin{equation}
c_3 = -\frac{4 n_{h,4}^{(1)}(d = 1) + 4 (2 n_{h,4}^{(2)}(d = \sqrt{2}) + 4 n_{h,4}^{(2)}(d = 2))}{2(v_{\rm rms}^0/t)^2} = c_{3}^{(0)} + c_{3}^{(J)} \cdot \left(\frac{J}{t}\right)^2 \simeq 0.172 + 0.136\cdot \left(\frac{J}{t}\right)^2, 
\end{equation}
gives the lowest order correction to the initial ballistic motion. Again, the numerical values are for the Heisenberg case in a 2D antiferromagnet. Note that this depends on the inverse interaction strength squared, $(J / t)^2$. In particular, we can use it to define the timescale for when the initial ballistic motion breaks down, by setting $1 - c_3 \cdot (t\cdot \tau_s)^2 = 1 / 2$. This yields Eq. (8) of the main text. 

\section{Exact initial expansion velocity}
In this section, we show that the exact initial velocity of a localized hole as described by the $t$-$J$ model has to be $\sqrt{z}t$, where $z$ is the number of nearest neighbors. \\

The $t$-$J$ model may be written concisely as
\begin{equation}
\hat{H} = \hat{H}_t + \hat{H}_J = -t \sum_{\braket{{\bf i}, {\bf j}}, \sigma} \left[ \tilde{c}^\dagger_{{\bf i},\sigma} \tilde{c}_{{\bf j},\sigma} + {\rm H.c.} \right] + J \sum_{\braket{{\bf i}, {\bf j}}} \!\left[ \hat{S}^{z}_{\bf i} \hat{S}^{z}_{\bf j} \!+\! \frac{\alpha}{2}\!\left(\hat{S}^{+}_{\bf i} \hat{S}^{-}_{\bf j} \!+\! \hat{S}^{-}_{\bf i} \hat{S}^{+}_{\bf j}\right) \!-\! \frac{\hat{n}_{\bf i} \hat{n}_{\bf j}}{4} \right],
\label{eq.H_t_J}
\end{equation}
where $\tilde{c}_{\bf j,\sigma} = \hat{c}_{\bf j,\sigma}(1 - \hat{n}_{\bf j})$ is the restrained fermionic annihilation operator. Since the hole is assumed to be initialized at the origin, $d = 0$, the initial state can be written as
\begin{equation}
\ket{\Psi(0)} = \ket{0}_{\bf 0}\left[U_{\boldsymbol\delta}\ket{\uparrow}_{\boldsymbol \delta}\ket{U_{\boldsymbol \delta}} + D_{\boldsymbol \delta}\ket{\downarrow}_{\boldsymbol \delta}\ket{D_{\boldsymbol \delta}} \right],
\label{eq.exact_initial_state}
\end{equation}
where the subscript is the site index. $\ket{0}_{\bf 0}$, thus, indicates a hole at the origin, ${\bf d} = {\bf 0}$. Also, $\boldsymbol\delta$ is the distance vector to one of the $z$ nearest neighbors, and $U_{\boldsymbol\delta}, D_{\boldsymbol\delta}$ indicate the respective amplitudes of the nearest neighbor to be spin-$\uparrow$ and -$\downarrow$. Hence, $|U_{\boldsymbol\delta}|^2 + |D_{\boldsymbol\delta}|^2 = 1$. We wish to compute the rms distance to linear order in time, $\tau$. To do so, the time evolution operator, $\hat{U}(\tau) = \te^{-i\hat{H}\tau}$ may be expanded to linear order in $\hat{H}$, and only $\hat{H}_t$ has to be retained. Hence, $\ket{\Psi}(\tau) = \hat{U}(\tau)\ket{\Psi(0)} \simeq [1 - i\hat{H}\tau]\ket{\Psi(0)}$, resulting in (all equal signs are to linear order in $\tau$)
\begin{align}
d_{\rm rms}^2 &= \sum_{\bf d} d^2 n_h(\bd, \tau) = \sum_{\boldsymbol\delta} n_h(\boldsymbol\delta,\tau) \nn \\
&= \sum_{\boldsymbol\delta} \bra{\Psi(0)}[1 + i\hat{H}_t\tau][1 - \sum_{\sigma}\hat{c}^\dagger_{{\boldsymbol\delta},\sigma}\hat{c}_{{\boldsymbol\delta},\sigma}] [1 - i\hat{H}_t\tau]\ket{\Psi(0)} = \tau^2 \sum_{\boldsymbol\delta} \bra{\Psi(0)}\hat{H}_t [1 - \sum_{\sigma}\hat{c}^\dagger_{{\boldsymbol\delta},\sigma}\hat{c}_{{\boldsymbol\delta},\sigma}] \hat{H}_t\ket{\Psi(0)}. 
\end{align}
Now, we write $\hat{H}_t = \sum_{\boldsymbol\delta} \hat{H}_t(\boldsymbol \delta) + \Delta H_t$, where $\hat{H}_t(\boldsymbol \delta)$ describes the hopping to and from the nearest neighbor at $\boldsymbol \delta$, wheras $\Delta H_t$ describe all the remaining hoppings in the rest of the lattice. From this and Eq. \eqref{eq.exact_initial_state}, it follows that
\begin{equation}
\hat{H}_t\ket{\Psi(0)} = t\sum_{\boldsymbol\delta'}\left[U_{\boldsymbol\delta'} \ket{\uparrow}_{\bf 0} \ket{0}_{\boldsymbol \delta'}\ket{U_{\boldsymbol \delta'}} + D_{\boldsymbol \delta'} \ket{\downarrow}_{\bf 0} \ket{0}_{\boldsymbol \delta'}\ket{D_{\boldsymbol \delta'}} \right].
\end{equation}
The hole counting operator, $[1 - \sum_{\sigma}\hat{c}^\dagger_{{\boldsymbol\delta},\sigma}\hat{c}_{{\boldsymbol\delta},\sigma}]$ applied 
to this state, only yields a nonzero result if ${\boldsymbol\delta} = {\boldsymbol\delta}'$. Therefore, 

\begin{align}
d_{\rm rms}^2 &= \tau^2 \sum_{\boldsymbol\delta} \bra{\Psi(0)}\hat{H}_t [1 - \sum_{\sigma}\hat{c}^\dagger_{{\boldsymbol\delta},\sigma}\hat{c}_{{\boldsymbol\delta},\sigma}] \hat{H}_t\ket{\Psi(0)} \nn \\
&= (t\cdot\tau)^2 \sum_{\boldsymbol\delta} \left[U_{\boldsymbol\delta}^*\bra{\uparrow}_{\bf 0} \bra{0}_{\boldsymbol \delta}\bra{U_{\boldsymbol \delta}} + D_{\boldsymbol \delta}^* \bra{\downarrow}_{\bf 0} \bra{0}_{\boldsymbol \delta}\bra{D_{\boldsymbol \delta}} \right] \left[U_{\boldsymbol\delta'} \ket{\uparrow}_{\bf 0} \ket{0}_{\boldsymbol \delta'}\ket{U_{\boldsymbol \delta'}} + D_{\boldsymbol \delta'} \ket{\downarrow}_{\bf 0} \ket{0}_{\boldsymbol \delta'}\ket{D_{\boldsymbol \delta'}} \right] \nn \\
&= (t\cdot\tau)^2 \sum_{\boldsymbol\delta} \left[|U_{\boldsymbol\delta}|^2 + |D_{\boldsymbol \delta}|^2\right] = (t\cdot\tau)^2 \sum_{\boldsymbol\delta} 1 = z (t\cdot\tau)^2,
\end{align}
whereby $d_{\rm rms} = \sqrt{z}t \cdot \tau$ to linear order in $\tau$. Hence, the initial velocity is $\sqrt{z}t$. For a square 2D lattice, $z = 4$, we, thus get $v_{\rm rms}^0 = 2t$. 

\section{Asymptotic polaron dynamics}
In this section, we analyze the asymptotic dynamics of the expansion coefficients in $\ket{\Psi_\bp(\tau)}$
\begin{equation}
\ket{\Psi_\bp(\tau)} = C^{(0)}(\bp; \tau) \cdot \hat{h}^\dagger_\bp\ket{\AF} + \sum_{\bk_1} C^{(1)}(\bp,\bk_1; \tau) \cdot \hat{h}^\dagger_{\bp + \bk_1} \hat{b}^\dagger_{-\bk_1} \ket{\AF} + \dots
\label{eq.wave_function_time_domain}
\end{equation}
to all orders, and show that this is solely governed by polaron dynamics. Since $\ket{\Psi_\bp(\tau)}$ is the Fourier transform of $\ket{\Psi_\bp(\omega)} = \ket{\Psi^{\rm R}_\bp(\omega)} + \ket{\Psi^{\rm A}_\bp(\omega)}$, it follows that
\begin{equation}
C^{(n)}(\bp, \{\bk_j\}; \tau) = \int \frac{{\rm d}\omega}{2\pi} \te^{-i\omega\tau} \cdot 2{\rm Re}\left[ R^{(n)}(\bp, \{\bk_j\}; \omega) \right]. 
\label{eq.C_n_from_R_n}
\end{equation}
Here, we also use that $\ket{\Psi^{\rm A}_\bp(\omega)} = [\ket{\Psi^{\rm R}_\bp(\omega)}]^*$. Before we dive into the general case, let us look at the simplest case of order $n = 0$. Here, we get
\begin{equation}
C^{(0)}(\bp; \tau) = -\int \frac{{\rm d}\omega}{2\pi} \te^{-i\omega\tau} \cdot 2{\rm Im}\left[ G^{\rm R}(\bp, \omega) \right] = \int \frac{{\rm d}\omega}{2\pi} \te^{-i\omega\tau} \cdot A(\bp, \omega),
\label{eq.C_n_from_R_A_n}
\end{equation}
with $A(\bp, \omega) = -2{\rm Im} [G^{\rm R}(\bp, \omega)]$ the hole spectral function. Importantly, this has a single well-defined peak at the quasiparticle energy $\varepsilon_\bp = \Sigma(\bp,\varepsilon_\bp)$, i.e. $A(\bp, \omega \sim \varepsilon_\bp) = 2\pi Z_\bp \cdot \delta(\omega - \varepsilon_\bp)$, with $Z_\bp$ the quasiparticle residue. At long times, this pole is the only remaining contribution, and, therefore, 
\begin{equation}
C^{(0)}(\bp; \tau) = \int \frac{{\rm d}\omega}{2\pi} \te^{-i\omega\tau} \cdot A(\bp, \omega) \to Z_\bp \cdot \te^{-i\varepsilon_\bp\tau}. 
\label{eq.C_0_asymptote}
\end{equation}
This is the principle line of reasoning, we will use in the following. In the general case, we utilize the recursion relation in Eq. \eqref{eq.recursion_relation} as well as the explicit coefficients in Eq. \eqref{eq.R_n_and_A_n_coefficients}. To make the calculations more concise in the following, we let $G^{\rm R}(\bK_n) = G^{\rm R}(\bK_n, \omega - \sum_{l = 1}^n \omega_{\bk_l})$, $A(\bK_n) = A(\bK_n, \omega - \sum_{l = 1}^n \omega_{\bk_l})$, and $C^{(n)} = C^{(n)}(\bp,\{\bk_j\}_1^n; \omega)$, $R^{(n)} = R^{(n)}(\bp,\{\bk_j\}_1^n; \omega)$. We, thus, obtain
\begin{align}
C^{(n)} &= 2\Re\left[R^{(n)}\right] = 2 g(\bK_{n-1},\bk_n) \Re\left[G^{\rm R}(\bK_n) R^{(n-1)} \right] \nn \\
&= 2 g(\bK_{n-1},\bk_n)\left\{\Re[G^{\rm R}(\bK_n)] \Re[ R^{(n-1)}] - \Im[G^{\rm R}(\bK_n)] \Im[ R^{(n-1)}]  \right\} \nn \\
&= g(\bK_{n-1},\bk_n)\left\{\Re[G^{\rm R}(\bK_n)] C^{(n-1)} + A(\bK_n) \Im[ R^{(n-1)}]  \right\} \nn \\
&= \Re\left[G^{\rm R}(\bK_0) \cdot \prod_{j=1}^{n-1} g(\bK_{j-1}, \bk_j) G^{\rm R}(\bK_j)\right] \cdot g(\bK_{n-1}, \bk_n) A(\bK_n) + g(\bK_{n-1},\bk_n)\Re[G^{\rm R}(\bK_n)] C^{(n-1)}.
\end{align}
In the second line, we utilize that the real part of a product of complex numbers is the product of the real parts minus the product of the imaginary parts. In the last line, we use that $\Im[R^{(n-1)}] = \Re[-i R^{(n-1)}] = \Re[G^{\rm R}(\bK_0) \cdot \prod_{j=1}^{n-1} g(\bK_{j-1}, \bk_j) G^{\rm R}(\bK_j)]$. Using this relation $n$ times yields (in which we now write out all terms explicitly)
\begin{align}
&C^{(n)}(\bp,\{\bk_j\}_1^n; \omega) = \prod_{j=1}^n g(\bK_{j-1}, \bk_j) \Bigg\{ A(\bp, \omega) \prod_{j=1}^n \Re[G^{\rm R}(\bK_j, \omega - \sum_{l = 1}^j \omega_{\bk_j})] \nn \\
 &+ \sum_{m = 1}^n A(\bK_m, \omega - \sum_{l = 1}^m \omega_{\bk_l}) \cdot \Re\left[G^{\rm R}(\bp,\omega) \prod_{j=1}^{m-1} G^{\rm R}(\bK_j, \omega - \sum_{l = 1}^j \omega_{\bk_l})\right] \cdot \prod_{j = m+1}^n \Re\left[ G^{\rm R}(\bK_j, \omega - \sum_{l = 1}^j \omega_{\bk_l}) \right] \Bigg\}.
 \label{eq.C_n_omega_general}
\end{align}
Importantly, this expression is now in the form of products of terms in which only a single quasiparticle pole appears. Specifically, terms like $\prod \Re[G^{\rm R}(\bK_j, \omega - \sum_{l = 1}^j \omega_{\bk_j})]$ contain no poles, as only the real part of the Green's functions appear. Also, the term $\Re[G^{\rm R}(\bp,\omega) \prod_{j=1}^{m-1} G^{\rm R}(\bK_j, \omega - \sum_{l = 1}^j \omega_{\bk_l})]$ have poles $\varepsilon_{\bK_j} + \sum_{l = 1}^j \omega_{\bk_l}$, but these all lie below the first pole of $A(\bK_m, \omega - \sum_{l = 1}^m \omega_{\bk_l})$ at $\varepsilon_{\bK_j} + \sum_{l = 1}^m \omega_{\bk_l}$. Therefore, only the latter pole contributes to the long time dynamics in these terms. In total, then, we may write down the explicitly asymptotic limit of the Fourier transform
\begin{align}
\!\!\!\!C^{(n)}(\bp,\{\bk_j\}_1^n; \tau) =&\, \prod_{j=1}^n g(\bK_{j-1}, \bk_j) \Bigg\{ Z_{\bp} \te^{-i\varepsilon_\bp\tau} \prod_{j=1}^n G^{\rm R}(\bK_j, \varepsilon_\bp - \sum_{l = 1}^j \omega_{\bk_j}) + \sum_{m = 1}^n \!Z_{\bK_m} \te^{-i(\varepsilon_{\bK_m} + \sum_{l=1}^m \omega_{\bk_l})\tau} \nn \\
&\times \Re\!\left[G^{\rm R}(\bp,\varepsilon_{\bK_m} \!+\! \sum_{l=1}^m \omega_{\bk_l}) \prod_{j=1}^{m-1}\! G^{\rm R}(\bK_j, \varepsilon_{\bK_m} \!+\!\! \sum_{l = j+1}^m \omega_{\bk_l})\right] \prod_{j = m+1}^n G^{\rm R}(\bK_j, \varepsilon_{\bK_m} \!-\!\! \sum_{l = m+1}^j \omega_{\bk_l}) \Bigg\}.\!
 \label{eq.C_n_tau_general}
\end{align}
Here, we use that in terms like $\Re[G^{\rm R}(\bK_j, \varepsilon_\bp - \sum_{l = 1}^j \omega_{\bk_j})]$, the Green's function is evaluated below the quasiparticle pole, and is, therefore, already real. Inserting this expression into Eq. \eqref{eq.wave_function_time_domain}, and grouping terms with respect to the overall factor of the residue, we see that
\begin{align}
\ket{\Psi_\bp(\tau)} =&\, Z_{\bp}\te^{-i\varepsilon_\bp \tau}\left[\hat{h}_\bp^\dagger + \sum_{n=1}^\infty \sum_{\{\bk_j\}_1^n} \hat{h}^\dagger_{\bK_n} \prod_{j=1}^n g(\bK_{j-1},\bk_j) G^{\rm R}(\bK_j, \varepsilon_\bp - \sum_{l=1}^j\omega_{\bk_l}) \hat{b}^\dagger_{-\bk_j} \right] \ket{\AF} \nn \\
&+ \sum_{\bk_1} g(\bp, \bk_1) \Re[G^{\rm R}(\bp,\varepsilon_{\bK_1} + \omega_{\bk_1})] Z_{\bK_1} \te^{-i(\varepsilon_{\bK_1} + \omega_{\bk_1})\tau} \hat{b}^\dagger_{-\bk_1} \nn \\
&\cdot \left[\hat{h}^\dagger_{\bK_1} + \sum_{n=2}^\infty\sum_{\{\bk_j\}_2^n} \hat{h}^\dagger_{\bK_n} \prod_{j=2}^n g(\bK_{j-1},\bk_j) G^{\rm R}(\bK_j, \varepsilon_{\bK_1} - \sum_{l=2}^j\omega_{\bk_l}) \hat{b}^\dagger_{-\bk_j} \right] \ket{\AF} + \dots 
\label{eq.wave_function_time_domain_asymptotics_1}
\end{align}
In this way, we have managed to write the asymptotic dynamics solely in terms of the magnetic polaron states \cite{Reiter1994,Ramsak1998,Nielsen2021}
\begin{equation}
\ket{\Psi_\bp^{\rm pol}} = \sqrt{Z_\bp} \left[\hat{h}_\bp^\dagger + \sum_{n=1}^\infty \sum_{\{\bk_j\}_1^n} \hat{h}^\dagger_{\bK_n} \prod_{j=1}^n g(\bK_{j-1},\bk_j) G^{\rm R}(\bK_j, \varepsilon_\bp - \sum_{l=1}^j\omega_{\bk_l}) \hat{b}^\dagger_{-\bk_j} \right] \ket{\AF}.
\label{eq.polaron_states}
\end{equation}
Explicitly, 
\begin{align}
&\ket{\Psi_\bp(\tau)} = \sqrt{Z_{\bp}}\te^{-i\varepsilon_\bp \tau}\ket{\Psi_{\bp}^{\rm pol}} \nn \\
&+ \sum_{n=1}^\infty \sum_{\{\bk_j\}_{j = 1}^n} \!\!\!\Re\left[\prod_{j=1}^n g(\bK_{j-1},\bk_j) G^{\rm R}(\bK_{j-1}, \varepsilon_{\bK_n} + \sum_{l=j}^n\omega_{\bk_l}) \right] \cdot \sqrt{Z_{\bK_n}} \te^{-i(\varepsilon_{\bK_n} + \sum_{l=1}^n \omega_{\bk_l})\tau} \cdot \prod_{j=1}^n \hat{b}^\dagger_{-\bk_j} \ket{\Psi_{\bK_n}^{\rm pol}}.
\label{eq.wave_function_time_domain_asymptotics_2}
\end{align}
This shows that the long time behavior is determined solely by magnetic polarons with $0, 1, \dots$ spin waves on top.

\section{Perturbative limit}
\begin{figure}[t!]
\begin{center}
\includegraphics[width=0.5\columnwidth]{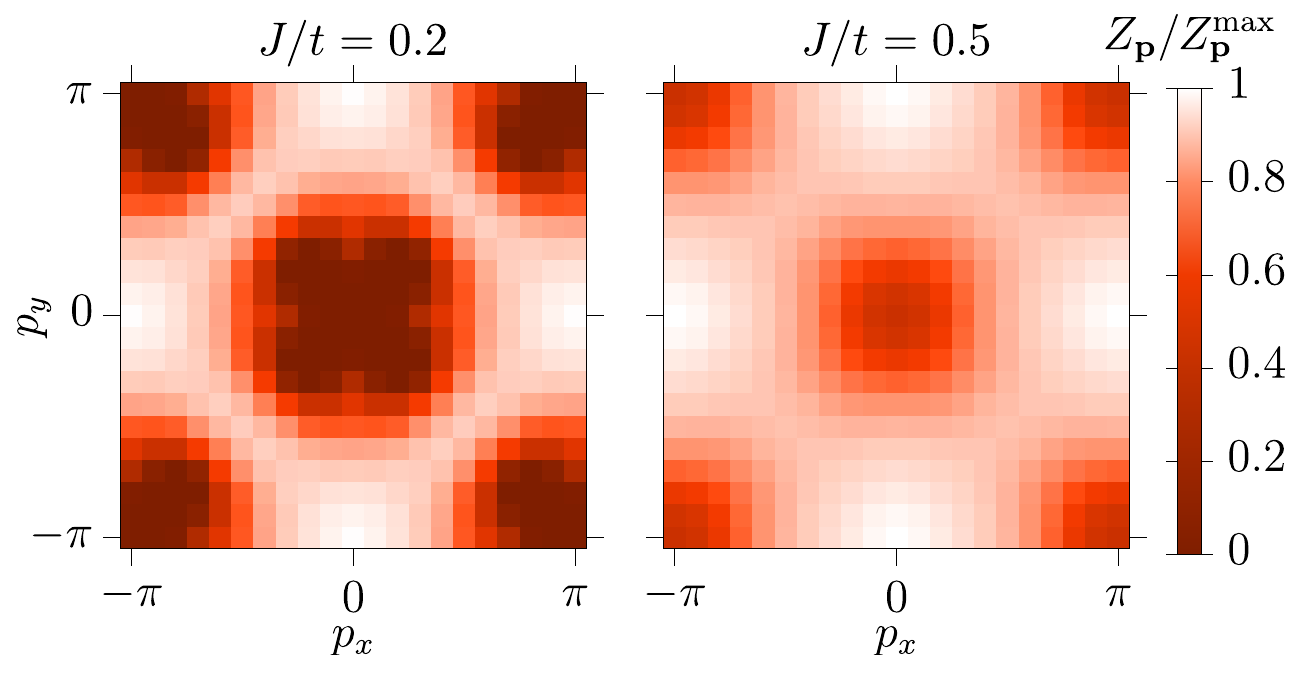}
\end{center}
\vspace{-0.7cm}
\caption{Quasiparticle residue for two indicated interactions strengths. This has been normalized to the maximal residue at each interaction strength to highlight differences in the residue across the Brillouin zone. For the strongly interacting case (left), the residue becomes vanishingly small in the center and corners of the Brillouin zone, constituting a clear qualitative difference to the behavior at moderate interactions (right). }
\label{fig.residue} 
\end{figure}

In this section, we calculate the polaron energy bandwidth $w_{\rm pol}$ and asymptotic expansion velocity $v_{\rm rms}^\infty$ in the perturbative limit to order $t^2 / J$. First, we calculate the polaron energy to order $t^2 / J$ by using the self-energy $\Sigma(\bp,\omega)$ to lowest order in the interaction [See Eq. (4) of the main text]
\begin{equation}
\varepsilon_\bp^{(0)} = \Sigma^{(0)}(\bp, 0) = -\sum_\bk \frac{g^2(\bp,\bk)}{\omega_\bk} = -\frac{t^2}{J} \sum_\bk \frac{\tilde{g}^2(\bp,\bk)}{\tilde{\omega}_\bk}.
\label{eq.perturbative_energy}
\end{equation}
Here, $\tilde{g} = g / t$ and $\tilde{\omega}_\bk = \omega_\bk / J$. From this, it is straightforward to get the perturbative polaron bandwidth as
\begin{equation}
w_{\rm pol}^{(0)} = \varepsilon^{(0)}_{\bf 0} - \varepsilon^{(0)}_{(\pi / 2, \pi / 2)} = c_w(\alpha) \frac{t^2}{J}. 
\label{eq.polaron_bandwidth_perturbative}
\end{equation}
Numerically, we determine $c_w(\alpha\to 1) \simeq 1.97$ for a square lattice of size $20\times 20$. To calculate the asymptotic expansion velocity, we need the mean-square distance
\begin{equation}
d_{\rm rms}^2 = \sum_{\bd} d^2 \bra{\Psi(\tau)} \hat{h}^\dagger_{\bd} \hat{h}_\bd \ket{\Psi(\tau)} = \sum_{\bp} \bra{\Psi(\tau)}\hat{h}^\dagger_{\bd}(-\nabla_\bp^2 ) \hat{h}_\bp \ket{\Psi(\tau)}. 
\end{equation}
From the previous section, we get the long time behavior of $\ket{\Psi_\bp(\tau)}$ up to first order in spin waves
\begin{align}
\ket{\Psi_\bp(\tau)} =& \, Z_\bp \te^{-i\varepsilon_\bp\tau} \hat{h}^\dagger_\bp\ket{\AF} \nn \\
&+ \sum_\bk g(\bp,\bk)\left\{ Z_\bp\te^{-i\varepsilon_\bp\tau} G^{\rm R}(\bp + \bk, \varepsilon_\bp - \omega_\bk) + Z_{\bp + \bk} \te^{-i(\varepsilon_{\bp + \bk} + \omega_\bk)\tau} \Re\left[ G^{\rm R}(\bp, \varepsilon_{\bp + \bk} + \omega_\bk) \right] \right\} \hat{h}^\dagger_{\bp + \bk} \hat{b}^\dagger_{-\bk}\ket{\AF}. 
\label{eq.Psi_p_asymptote_first_order}
\end{align}
In principle, we, then, need to calculate the double derivative of this full expression with respect to momentum. However, at long times, only terms in which the phases are derived with respect to time contribute significantly, since they will scale as $\tau^2$. The lowest order result for the mean-square distance at long times is, therefore, simply 
\begin{equation}
d_{\rm rms}^2 \to \tau^2 \frac{1}{N}\sum_\bp (\nabla_\bp \varepsilon_\bp^{(0)})^2 = c_v^2(\alpha)  \cdot \left(\frac{t^2}{J}\right)^2\tau^2, 
\end{equation}
replacing $\varepsilon_\bp$ with $\varepsilon_\bp^{(0)}$ in Eq. \eqref{eq.Psi_p_asymptote_first_order}. Hence, $v_{\rm rms}^\infty = c_v \cdot t^2 / J$. Using Eq. \eqref{eq.perturbative_energy}, we get
\begin{align}
c_v^2(\alpha) = \frac{2^4}{N^2}\sum_{\bp,\bq,\bk} \frac{\tilde{g}(\bp,\bq) \tilde{g}(\bp,\bk)}{\tilde{\omega}_\bk \tilde{\omega}_\bq}\big\{& \left[u_\bk \sin(p_x + k_x) - v_\bk \sin(p_x)\right]\left[u_\bq \sin(p_x + q_x) - v_\bq \sin(p_x)\right] \nn \\
+&\left[u_\bk \sin(p_y + k_y) - v_\bk \sin(p_y)\right]\left[u_\bq \sin(p_y + q_y) - v_\bq \sin(p_y)\right] \big\}.
\label{eq.c_v_squared}
\end{align}
Numerically, we, hereby, determine $c_v(\alpha \to 1) = 0.81$ for a square lattice of size $20\times 20$. This yields the perturbative limit of the ratio
\begin{equation}
\frac{v_{\rm rms}^\infty}{w_{\rm pol}} = \frac{c_v}{c_w} \simeq 0.41,
\label{eq.v_rms_to_w_pol_perturbative}
\end{equation}
where the numerical value is in the Heisenberg limit of $\alpha = 1$. We have checked the finite size dependency of this ratio, which seems to be within $0.1\%$. As mentioned in connection with Fig. 4 of the main text, the relation in Eq. \eqref{eq.v_rms_to_w_pol_perturbative} remains roughly constant until $J \lesssim 0.4t$, where $v_{\rm rms}^{\infty}/w_{\rm pol}$ suddenly starts to drop. We associated this with a qualitative change in the behavior of the quasiparticle residue throughout the Brillouin zone. We elucidate this in more detail in \ref{fig.residue}, in which the residue is plotted as a heat map for two indicated interactions strengths. In the strongly interacting case of $J = 0.2t$, the residue becomes vanishingly small in the center and corners of the Brillouin zone, whereby these states cannot efficiently contribute to the long-time dynamics of the moving hole. 

\section{Energy and lifetime of lowest string excitation}
\begin{figure}[t!]
\begin{center}
\includegraphics[width=0.95\columnwidth]{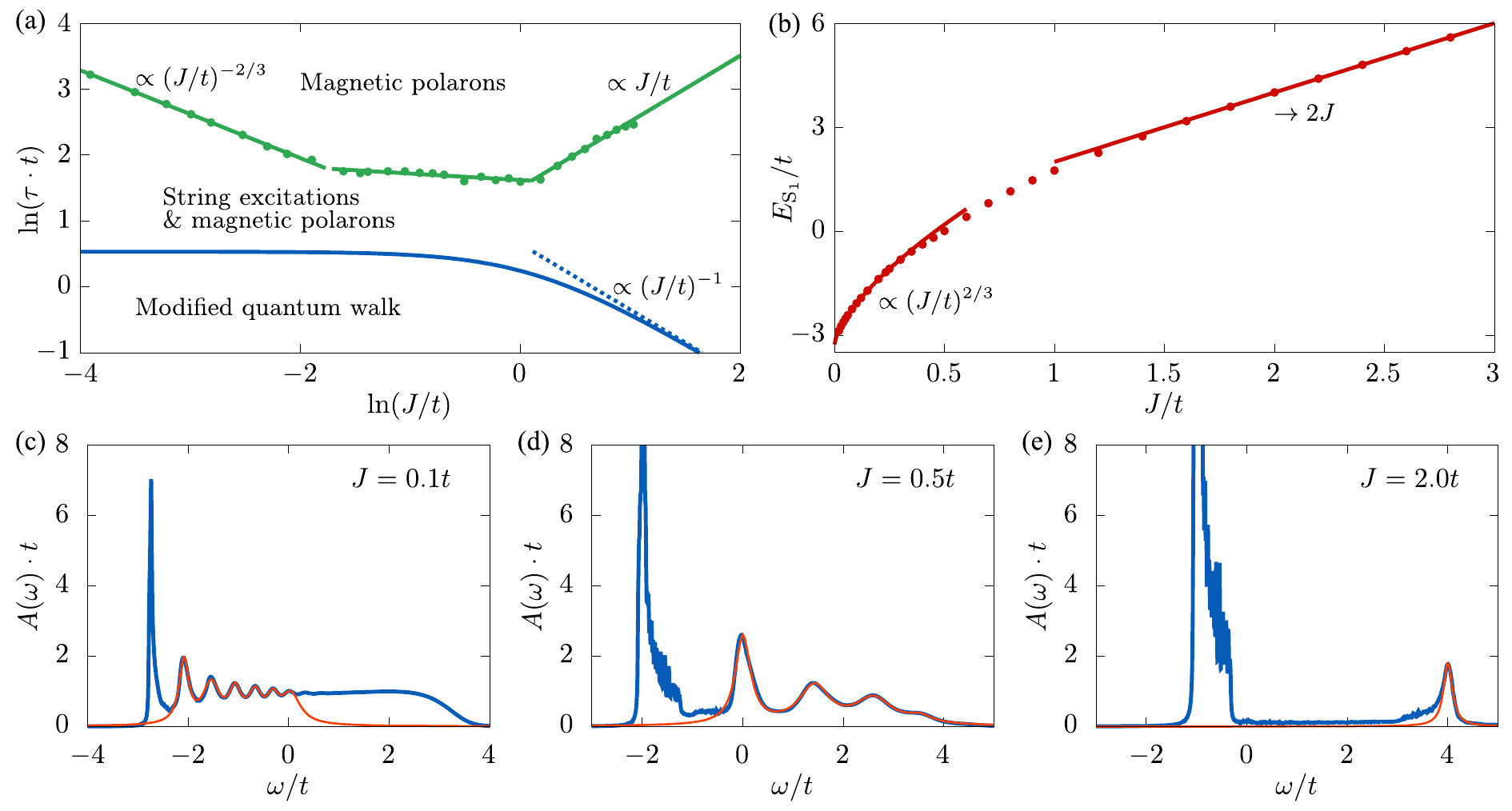}
\end{center}
\vspace{-0.7cm}
\caption{(a) Extracted lifetime [green points] along with power-law fits (green solid lines) of the lowest string excitation, ${\rm S}_1$. For better comparison with Fig. 1(c) of the main text, we also show $\tau_s$ [Eq. (8) of the main text]. (b) Associated energy of the lowest energy string excitation (red points) along with power-law fits in the strong ($J \ll t$) and weak ($J \gg t$) coupling limits. (c)--(e) Exemplary multi-Lorentzian fits for three indicated interactions strengths.}
\label{fig.width_and_energy_string} 
\end{figure}

In this section, we show our numerical evidence for the presence of the lowest string excitation, see Figs. 3(e) and 3(f) of the main text. To extract these, we perform a Lorentzian fit
\begin{equation}
L(Z, E, \Gamma; \omega) = 2Z\frac{1 / \tau}{(\omega - E)^2 + (1/\tau)^2},  
\end{equation}
with the residue $Z$, the energy $E$ and the lifetime $\tau$ as fitting parameters. In Fig. \ref{fig.width_and_energy_string}, we show the results for $E$ and $\tau$ across a range of interactions strengths $J / t$, with exemplary multi-Lorentzian fits shown in Figs. \ref{fig.width_and_energy_string}(c)-\ref{fig.width_and_energy_string}(e). \vspace{0.5cm}

\bibliographystyle{apsrev4-1}
\bibliography{ref_magnetic_polaron}